# IGM Emission Observations with the Cosmic Web Imager: II. Discovery of Extended, Kinematically-Linked Emission around SSA22 Lyα Blob 2

D. Christopher Martin[1], Daphne Chang[1,2], Matt Matuszewski[1], Patrick Morrissey[1], Shahin Rahman[1], Anna Moore[3], Charles C. Steidel[4], Yuichi Matsuda[4]

[1]Cahill Center for Astrophysics, California Institute of Technology, 1216 East California Boulevard, Mail Code 278-17, Pasadena, California 91125, USA. [2]Deceased. [3]Caltech Optical Observatories, Cahill Center for Astrophysics, California Institute of Technology, 1216 East California Boulevard, Mail Code 11-17, Pasadena, California 91125, USA. [4]Cahill Center for Astrophysics, California Institute of Technology, 1216 East California Boulevard, Mail Code 249-17, Pasadena, California 91125, USA

**Abstract**

The intergalactic medium (IGM) is the dominant reservoir of baryons, delineates the large scale structure of the universe at low to moderate overdensities, and provides gas from which galaxies form and evolve. Simulations of a Cold Dark Matter (CDM) dominated universe predict that the IGM is distributed in a cosmic web of filaments, and that galaxies should form along and at the intersections of these filaments (Bond, Kofman, & Pogosyan 1994; Miralda-Escude et al. 1996). While observations of QSO absorption lines and the large-scale distribution of galaxies have confirmed the CDM paradigm, the cosmic web of IGM has never been confirmed by direct imaging. Here we report our observation of the Lyα blob-2 (LAB2) in SSA22, with the Cosmic Web Imager. This is an integral field spectrograph optimized for low surface brightness, extended emission. With 22 hours of total on- and off-source exposure, CWI has revealed that LAB2 has extended Lyα emission which is organized into azimuthal zones consistent with filaments. We perform numerous tests with simulations and the data to secure the robustness of this result, which relies on data with modest signal-to-noise ratio. We have developed a smoothing algorithm that permits visualization of data cube slices along image or spectral-image planes. With both raw and smoothed data cubes we demonstrate that the filaments are kinematically associated with LAB2 and display double-peaked profiles characteristic of optically thick Lyα emission. The flux is 10-20 times brighter than expected for the average emission from the IGM but is consistent with boosted fluorescence from a buried QSO or gravitation cooling radiation. Using simple emission models we infer a baryon mass in the filaments of at least $1-4\times10^{11} M_{\odot}$, and the dark halo mass is at least $2\times10^{12} M_{\odot}$. The spatial-kinematic morphology is more consistent with inflow from the cosmic web than outflow from LAB2, although an outflow feature maybe present at one azimuth. LAB2 and the surrounding gas have significant and coaligned angular momentum, strengthening the case for their association.

## 1. Introduction

Theories of cosmological structure formation in a Cold Dark Matter (CDM) dominated universe predict that the matter in the intergalactic medium (IGM) and the galaxies formed from it are organized in a "cosmic web" of walls and filaments (Davis et al. 1985; Frenk et al. 1985; Bond, et al. 1994; Miralda-Escude, et al. 1996) – with structures spanning scales ranging from about one megaparsec to a few gigaparsecs (Mpc, Gpc, comoving). The intersections of sheets and filaments mark regions of high overdensity in which galaxies form and thrive, while the vast spaces between them are devoid of matter in comparison. Galaxy redshift surveys uncovered this morphology on large scales several decades ago (de Lapparent, Geller, & Huchra 1986). The gas associated with the filaments has been probed for nearly half a century via absorption features in the spectra of high redhsift quasars (Gunn & Peterson 1965; Lynds 1971). A comprehensive review of the derived IGM properties is given in Meiksin (2009). These investigations have revealed the evolution of the IGM, its enrichment with metals and association with galaxies. For instance, the spectral density of the Lyman alpha (Lyα) forest lines at redshift z ~ 3 indicates that nearly all baryons in the universe at that time were in a relatively cool phase ($T \sim 10^4$K) (Rauch et al. 1997). At lower redshifts, near the current epoch, the Lyα forest thins out considerably with a large fraction of the cool gas being shock heated as it undergoes further gravitational collapse within the filaments, giving rise to a warm-hot intergalactic medium (WHIM; Bregman 2007). Some of this gas accretes onto the virialized dark matter halos coalescing onto galaxies. Numerical simulations of structure formation are consistent with this picture, suggesting that cosmic web forming baryonic matter traces the underlying dark matter well (Cen et al. 1994; Zhang, Anninos, & Norman 1995; Hernquist et al. 1996; Fukugita & Peebles 2004; Cen & Ostriker 2006; Prochaska et al. 2009). The formation and evolution of galaxies is expected to be intimately connected to the IGM. Recent modeling indicates that galaxies may acquire a large fraction of their stellar fuel via streams of cold gas flowing along IGM filaments. This gas can fuel rapid periods of star formation and could explain the abundance of starburst galaxies observed around z ~ 2 (Birnboim & Dekel 2003; Kereš et al. 2005; Faucher-Giguère, Kereš, & Ma 2011; van de Voort et al. 2011ab). The computer models are being continually refined, however, resulting in modifications of the predictions of the relative impact of "cold" and "hot" mode gas accretion (Nelson et al. 2013). Observations of the morphology and kinematics of the IGM around galaxies and its correlation and co-evolution with them are necessary to vet the computer models and to further test our understanding of cosmic structure formation and galaxy assembly and evolution (Kirkman & Tytler 2008; Faucher-Giguère & Kereš 2011; Fumagalli et al. 2011). While QSO absorption sightlines have yielded volumes of information about the co-evolution of IGM gas and galaxies, the information is limited in nature due to the sparseness of suitable background continuum sources on the sky and is, largely, statistical. Probing the full three-dimensional structure of specific systems requires another approach – either relying on a different population of background sources (more numerous or extended; nominally higher redshift galaxies) or measuring emission from the systems of interest. The former is likely not feasible until the 30 meter class telescopes come online, while the latter is within the capabilities of current instrumentation, particularly in the case of systems developing in regions of high overdensity.

Emission in strong UV resonance lines, particularly HI Lyα 1216Å, is predicted to delineate the cosmic web (Hogan & Weymann 1987; Binette et al. 1993; Gould & Weinberg 1996a; Furlanetto et al. 2003,2005; Cantalupo et al. 2005; Kollmeier et al. 2010). For the vast majority of gas within the cosmic web the emission is expected to be faint (~$10^{-20}$ erg/s/$cm^2$/$arcsec^2$), as the only source of ionizing radiation is the relatively weak metagalactic UV background. The presence of strong ionizing sources, such as AGN, quasars (Haiman & Rees 2001; Cantalupo, et al. 2005), or even regions of star formation within a galaxy, will boost (by up to several orders of magnitude) the Lyα signal emanating from the surrounding gas. Depending on the particular geometry and gas column density, the diffuse emission may be limited to the galaxy halo or it may extend well into the IGM. The nature of the emission may vary from resonantly scattering Lyα photons into our line of sight in cases when the gas is optically thin to Lyα fluorescence when the clouds are self-shielding. Collisional excitation due to gravitational infall, as is expected to occur in the case of cold accretion flows (Fardal et al. 2001; Förster Schreiber et al. 2006; Faucher-Giguère et al. 2010) may also lead to Lyα emission with surface brightness that may rival the signal stemming from fluorescence induced by a strong ionizing source (Dijkstra, Haiman, & Spaans 2006). Finally, the possibility that the diffuse emission is driven by mechanical feedback from the galaxy, in particular superwinds, should not be discounted (Taniguchi & Shioya 2000; Ohyama et al. 2003; Mori, Umemura, & Ferrara 2004).

Significant effort is being exerted to detect and characterize the IGM in emission. Deep narrowband observations (e.g., Fynbo, Møller, & Warren 1999; Steidel et al. 2000; Hayashino et al. 2004; Nilsson et al. 2006; Matsuda et al. 2011) have imaged numerous Lyα blobs in overdense regions while Cantalupo, Lilly, and Haehnelt (2012) report a detection of a rich field of Lyα sources around a z~2.4 quasar. Some of the observed nebulae are known to harbor obscured active galactic nuclei (AGN; e.g., Basu-Zych & Scharf 2004; Chapman et al. 2004; Barrio et al. 2008; Overzier et al. 2013), while for others the energy source is less clear, with dissipation of gravitational potential energy being a possible culprit (e.g., Nilsson, et al. 2006). Various techniques, both imaging and spectroscopic, have been used to search for Lyα halos around galaxies and quasars (e.g., Hu & Cowie 1987; Francis & Bland-Hawthorn 2004; Hennawi et al. 2009; Steidel et al. 2011; North et al. 2012). Long- and multi-slit spectroscopic investigations have resulted in detections of a number of Lyα emitters in the vicinity of quasars (Rauch et al. 2008). There have also been reports of emission associated with damped Lyα systems near QSOs, although the observed flux is likely to originate from either the damped Lyα (DLA) host galaxy or the background quasar (Adelberger et al. 2006; Hennawi, et al. 2009). More recently, spectroscopic surveys have yielded intriguing observations of Lyα emission from the IGM around quasars and galaxies exhibiting possible filamentary nature and a intricate velocity structure (Hennawi & Prochaska 2013; Rauch et al. 2013). The evidence appears strong that the IGM around galaxies and quasars is rich in gas that manifests a complex morphology, kinematic structure, and interacts extensively with the central object. Unfortunately, the above-outlined observations and techniques do not result in a full picture of the individual systems being studied, yielding either incomplete spatial or limited spectral information. Integral field spectroscopy, however, combines moderate resolution spectral content with spatial data over a compact field that is well matched in shape and size to the circum-galactic environments. As such, it is a great tool for probing galaxies and their neighborhoods on scales relevant to galaxy assembly and feedback. We have initiated a program to study faint emission around galaxies, quasars, and Lyα nebulae with the Palomar Cosmic Web Imager, an integral field spectrograph for the Hale Telescope at Palomar Observatory built specifically to detect and map dim and diffuse emission.

In this paper we report the detection of extended, filamentary emission intersecting at Lyα blob-2 (LAB2) in the known overdense SSA22 field at $z \approx 3.09$ (Steidel et al. 1998; Steidel, et al. 2000). This is an extensively studied object with multiband imaging data available (e.g., Steidel, et al. 2000; Chapman et al. 2001; Basu-Zych & Scharf 2004; Geach et al. 2007; Webb et al. 2009; Nestor et al. 2011). The blob is, roughly, 15 × 10 arc seconds in size and has a Lyα luminosity of nearly $10^{44}$ erg/s. It is known to harbor an enshrouded AGN (Basu-Zych & Scharf 2004). Spectroscopic investigations of LAB2 (Steidel, et al. 2000; Wilman et al. 2005; Matsuda et al. 2006) have revealed broad (300 – 400 km/s) Lyα emission lines with an absorption-like feature at line center. The detected velocity shear across the blob reaches nearly 2000 km/s and is indicative of the complexity of the system. We have used PCWI to co-add a 50 × 50 arc second field around LAB2 in order to map the morphology and kinematics of the surrounding gas. In §2 we describe the instrument, data acquisition, reduction, and analysis required to generate a sky-subtracted unsmoothed data cube. In §3 we show that we detect extended emission around LAB2 that is kinematically associated with LAB2. In §4 we debate alternative explanations for the extended emission. In §5 we discuss the morphology, kinematics, and energetics of the extended emission, and place these results in the context of current models for emission from the IGM and gas flows into forming galaxies. We summarize the main findings in §6. We have also included an Appendix that reviews the adaptive smoothing algorithm as well as a series of tests and simulations designed to demonstrate the robustness of the algorithm and the extended emission detections around LAB2. All results assume a WMAP-7 cosmology (Komatsu et al. 2011) and, when given, linear dimensions are in proper coordinates.

## 2. Cosmic Web Imager Observations

### 2.1 The Palomar Cosmic Web Imager

We have constructed an integral field spectrograph called the Palomar Cosmic Web Imager (CWI), that is designed to search for, map, and characterize IGM emission and other low surface brightness phenomena (Matuszewski et al. 2010). It uses a 40" × 60" reflective image slicer with twenty-four 40" × 2.5" slices. The spectrograph has a high dispersion volume phase holographic grating covering 4500-5400Å with an instantaneous bandwidth of 400Å (without nod and shuffle, see below). The spectrograph attains a resolution of $\Delta\lambda$~1Å and a peak efficiency of ~10% at 5000Å including the telescope and atmosphere. CWI is mounted at the Cassegrain focus of the Hale 5 meter telescope on Mt. Palomar. As a spectroscopic imager it is possible to form individual 1Å wide images of the sky. The imaging resolution, while limited by the 2.5" slicer sampling, can be effectively improved to ~1.3" by dithering the field between individual exposures. A description of the instrument, general observing approach, and data analysis methodology is given in Paper I (Martin et al. 2013).

### 2.2 Science Data Observations

Since it is likely that IGM emission is brightest in the highest overdensity regions of the universe, we observed Ly$\alpha$ blob-2 (LAB2) in the well known SSA22 (Steidel, et al. 1998; Steidel, et al. 2000) overdensity field at z=3.09. A smoothed, stretched narrow-band image of LAB2 around the Ly$\alpha$ line is shown in Figure 2 (Hayashino, et al. 2004; Matsuda et al. 2004). The science data cube was constructed from observations during three separate runs. In the first run a single background field was used. In the second and third runs nod-and-shuffle was implemented. The three runs were performed at three different position angles (0°, 180°, and 270°) to dilute the impact instrumental effects. A total exposure of 11 hours on source and 11 hours on sky was obtained over the three runs.

#### 2.2.1 Run 1

We obtained ten 20-minute exposures on source (LAB2), on 2-3 October 2010, centered on position ($\alpha$,$\delta$)=(22:17:39, +00:13:27) and interleaved with eleven 20-minute exposures at a nearby sky position ($\alpha$,$\delta$)=(22:17:32, +00:14:30), a location with no known z=3.1 structures or bright objects. We also obtained calibration images using lamps throughout the observing period.

The recorded 2D spectra are sliced, rectified, spatially alinged using pinhole mask calibration images, and wavelength-shifted to compensate for <1Å of total flexure using sky emission lines. Exposure maps are generated by processing flat-field images (calibration and twilight flats) in a similar fashion. The result is a set of data cubes (RA, DEC, and λ) for each exposure, sampled at (0.292", 0.292", 1Å) covering 4800-5300Å. Slices were oriented east-west, with position angle 0°.

Sky levels typically varied smoothly by 0.5-2% between the 20-minute source and sky exposures, and emission lines in the LAB2 velocity range (notably sodium broad [several tens of Å] and several narrow lines over 4870-4984Å) varied most strongly. Sky levels were ~1-2 × $10^5$ LU/Å (ph $cm^{-2}$ $s^{-1}$ $sr^{-1}$ $Å^{-1}$)[1], with a contribution of 25-50% from broad NaI and ~10% from narrow NaI lines. Our emission line detections are at the level of ~1% of sky. Sky exposures for Run 1 were interpolated in time and wavelength to correspond to the sky level in the interleaving source exposures. The corrected sky cubes were then subtracted from the source cubes and the difference cubes were divided by the exposure maps to generate the flux cubes.

### 2.2.2 Run 2

For Run 2, we obtained a total of 6 hours on-source and 6 hours off-source exposure centered on LAB2 on 29 September – 1 October 2011. Individual exposures were obtained using a nod-and-shuffle technique (Cuillandre et al. 1994; Sembach & Tonry 1996; Glazebrook & Bland-Hawthorn 2001). We used the central 1/3 of the CCD for recording the spectrum, and masked the outer 2/3 of the detector (restricting the spectral bandpass to ~150Å and the effective common bandpass to ~85Å given the image slicer offset brick wall pattern [see Figure 3] and slit curvature). Each frame was created by interleaving an integral number, N, of on-target telescope pointings of duration t with N+1 background pointings, the first and last of length t/2, the remainder, t. This results in separate source and background images being built up nearly contemporaneously on the CCD, each equivalent to an (N × t) exposure. This observing procedure starts with the telescope pointed at the background field. The shutter is opened and data is collected for time t/2. The shutter is closed and the charge on the CCD is shuffled by 670 pixels, approximately 1/3 of the CCD, and stored under a physical mask. The telescope is nodded to the target field. The shutter is re-opened, this time for time t, and source data is collected. Once the shutter closes, the charge is shifted by -670 pixels so that the newly collected data is blocked, while the previously acquired background image is uncovered and the telescope is nodded back to the background field. A background exposure of length t is taken, followed by a source exposure of length t, etc. The process is repeated until the desired source integration time ($N \times t$) is reached, at that point the final book-end t/2 background data is collected and the CCD is read out. A typical 40 minute exposure (t = 120 s, N = 10; 20 minutes source + 20 minutes background) takes approximately 50 minutes of wall-clock time, including CCD read-out. The purpose of employing the nod-and-shuffle method is to sample the sky as frequently as possible and to do so through the exact same optical path and using the same detector pixels as those used for the object. This improves sky subtraction precision, reduces the contribution of detector read noise by decreasing the number of CCD read-outs, and limits the impact of instrument systematics. To further reduce the impact of the $3e^-$ (rms) read noise, the detector is binned 2 by 2.

[1] 1 LU = 1 Line Unit = 1 ph $cm^{-2}$ $s^{-1}$ $sr^{-1}$. For continuum spectra we use LU/Å. 1000 LU corresponds to ~$10^{-19}$ erg $cm^{-2}$ $s^{-1}$ $arcsec^{-2}$ at 5000Å.

During Run 2 a total of 18 nod-and-shuffle exposures was obtained over three observing nights, totaling 6 hours of integration time on source and an equal amount on background. The PCWI IFU was oriented with slices in the north-south direction at position angle 270°. Source pointings were dithered by ½ slit width west over 5 arcsec. They were alternately pointed 5 arcsec north $(\alpha,\delta)$=(22:17:39, +00:13:33) and 5 arcsec south $(\alpha,\delta)$=(22:17:39, +00:13:23) of LAB2 in order to produce good exposure over a 50 arcsec by 50 arcsec field of view and provide the advantage of further averaging over instrument features. Four half-slice east-west dithers were used to further average out slice-to-slice efficiency variations. Background pointings were dithered by 7.5 arcsec west each exposure for a total of 90 arcsec, in order to smooth out and minimize any local background features. This background area was chosen because it has minimal bright narrow-band or continuum sources.

In Figure 3 we show a single nod-and-shuffle frame, with source, background, and difference frames, prior to most pipeline processing. No subtraction residuals are apparent.

### 2.2.3 Run 3

During Run 3 over the nights of 23 and 24 October 2011, we obtained a total six nod-and-shuffle frames, totaling 2 hours of on-source and 2 hours off-source exposure centered on LAB2. The slices were oriented east-west, with position angle 180°. The observing method and data reduction was otherwise identical to Run 2.

For runs 2 and 3, sky levels were again ~1-2 × $10^5$ LU/Å with a contribution of 25-50% from broad Na and ~10% from narrow Na lines. , and sky levels typically varied smoothly and mostly linearly by 0.5-2% on 20 minute timescales or typicaly 0.1% on the 2 minute nod-and-shuffle cadence, leading to sky-subtraction errors significantly less than 300 LU/Å for individual exposures, and correspondingly smaller errors for the summed exposures.

### 2.3 Data Cube Generation

We obtained numerous calibration images using lamps throughout each observing period. The obtained 2D spectra are sliced, rectified, spatially aligned using pinhole and edge mask calibration images and arc lamp spectra, and wavelength-shifted to compensate for <1Å of total flexure using sky lines. The final individual run and 3-run mosaiced data cubes are generated using the astrometry based on the guide star camera, Additionally, the astrometry was checked using several moderately bright objects in the source fields. The reconstruction is good to about 0.5 arcsec (rms). Mosaiced data cubes were created for source and background. Source data cubes are rectified based on astrometry, and background data cubes were created using the same source data cube registration (so that individual source and background exposures are subtracted one-to-one). Compact source subtraction is discussed in §3.2 below.

Exposure maps are generated by processing normalized flat-field images (calibration and twilight flats) in a similar fashion. The result is a set of data cubes (RA, DEC, and λ) for each exposure, sampled at (0.292", 0.292", 1Å) covering 4900-5050Å. The difference cube is obtained by simply subtracting source and background cube. Because the nod-and-shuffle mask does not physically contact the CCD, a small amount of diffuse continuum light remains in the subtracted cube (<1%), which is easily subtracted with a low-order continuum fit.

**2.4 Statistical Consistency of Runs 1-3 and Co-addition**

We checked for statistical consistency by comparing slices from Runs 1-3. Figure 1 shows such a comparison. In general the three runs were consistent within predicted Poisson errors as discussed in the Figure caption. The principal remaining uncertainty is the possible presence of emission features at the local redshift in the background fields. The summed background field consists of a superposition of ~10 different background fields and even more dithers. All background fields were chosen to be distant from any low surface brightness features in the narrow-band images.

A coadded difference cube was generated by an exposure-weighted sum of the registered individual run data cubes. A total exposure of 11 hours on source and 11 hours on sky was obtained over the three runs at the three different position angles (0°, 180°, and 270°). The result is a sky- and source-subtracted data cube of ~50" × 50" × 1Å slices with 0.292" pixels, and a 1-sigma noise level of ~2200 LU in 5" × 5" × 4Å bins, or 550 LU/Å.

**3. Detection of Extended Lyα Emission Near the Systemic Velocity**

**3.1 Unsmoothed Data Cube: Comparison to Narrow-band Image**

A smoothed, stretched narrow-band image of LAB2 around the Lyα line is shown in Figure 2 (Hayashino, et al. 2004; Matsuda, et al. 2004), and displays evidence for extended emission. We superpose a radial/azimuthal grid for reference. The narrow-band *NB497* and broad-band (*B*, *V*) data were taken with Suprime-Cam on the 8.2 m Subaru Telescope (Miyazaki et al. 2002). The narrow-band filter, *NB497*, has the central wavelength (CW) of 4977Å and FWHM of 77Å to detect Lyα emission line at z = 3.06-3.13. The spatial variation of the CW and the FWHM of the *NB497* filter are less than 18Å (Δz=0.015) and 5Å (Δz=0.004), respectively. The raw data were reduced with SDFRED (Yagi et al. 2002; Ouchi et al. 2003). We made flat fielding using the median sky image and background sky subtraction adopting the mesh size parameter of 30" before combining the images. All the stacked images were calibrated using spectrophotometric standard stars (Massey et al. 1988; Oke 1990) and Landolt standard stars (Landolt 1992). The magnitudes were corrected for Galactic extinction of *E(B-V)=0.08*. The combined images were aligned and smoothed with Gaussian kernels to match their seeing sizes. The average stellar profile of the final images has FWHM of 1.0". We constructed a *BV* image [*BV~(2B+V)/3*] for the continuum at the same effective wavelength as the narrow-band filter and an emission line *NB-BV* image by subtracting the *BV* image from the *NB497* image. The limiting magnitudes (1σ) per square arcsecond are 28.8(*NB497*), 29.1(*BV*), and 28.8(*NB-BV*).

In Figure 4 we compare the narrow-band image, sky-subtracted but not continuum-subtracted, to a stacked CWI pseudo-narrow-band image created with a Gaussian filter, FWHM 77Å, both images displayed with a similar intensity scale. A radial grid is superposed to aid in cross-identification of emission features. The narrow-band image has a nominal Poisson noise level of 2.5 x $10^{-19}$ erg $cm^{-2}$ $s^{-1}$ $arcsec^{-2}$ (2500LU) for this smoothing. Because of the narrow-band filter width (77Å), typical sky (60,000LU/Å) corresponds to a total level of 5 x $10^6$ LU. A 0.1% error in sky subtraction would lead to a 5000LU error in the corrected narrow-band image. Sky-subtraction errors in the narrow-band images are introduced by a number of factors, including the spatial/spectral variation in the instrument response over the narrow vs. broad continuum filter, the variation in the sky spectrum over the two wavelength ranges, the impact of continuum source wings and scattered light, and the additional narrow-band objects present in the broader continuum image. Spatially extended structures are particularly subject to these errors. We estimate from variations in the zero level that typical systematic continuum-subtraction errors are ~0.2% of sky, or ~10 x $10^{-19}$ erg $cm^{-2}$ $s^{-1}$ $arcsec^{-2}$ (~10000LU). The CWI stacked image with the 5 arcsec smoothing has a similar Poisson error level (~7000-10000LU). Thus for both images typical noise amplitude will cause variations from blue to green color values on 5 arcsec scales, as is observed. Comparison of individual features detected in either image will not be one-to-one because of this noise level.

### 3.2 Unsmoothed Data Cube: Spatial and Spectral Properties

In order to explore the morphology of this emission and to isolate distinct cosmic structures we investigate slices near the LAB2 emission velocity range, noting that a single 1Å slice corresponds to a line-of-sight distance of 5.3 comoving Mpc (Bennett et al. 2003). We first show that extended emission around LAB2 is detected with high significance. Radially and azimuthally binned image slices are displayed in Figure 5 in six 4Å wavelength bins near the systemic velocity of LAB2. Extended emission is detected in some of these spatial bins in all of the 4Å slices, out to a radius of 30 arcsec. The final panel shows a 32Å bin around the wavelength 4982Å (as we discuss below the systemic velocity is 4981.7Å). While there is indication that the extensions are azimuthally confined, with concentrations in azimuth 0°-90°, 150°-240°, and 300°-360°, the modest signal-to-noise ratio and complex spatial and spectral morphology erects a challenge for simple interpretation.

We can also investigate spectra obtained in individual spatial bins. This is done in Figure 6. Again, even thought the spectra have modest signal-to-noise ratio, prominent emission lines are detected, with evidence in some cases for double-peaked profiles characteristic of optically thick Lyα emission (Neufeld 1990). Each row shows a fixed azimuthal bin with a succession of radial bins. There is a gradual evolution of the spectrum going outward from the blob, suggesting that the extended emission is kinematically related to the blob emission. However, the signal-to-noise ratio is modest, and it would be useful to have additional tools for examining the extended emission spectrally and spatially, for adaptively smoothing in a fashion that highlights the spectral and spatial morphology of the extended emission, and for relating the extended emission to that of the blob.

In general, even long exposure data from CWI targeting IGM emission is of modest signal-to-noise ratio. IGM emission models predict that line emission will be extended, probably filamentary, and may show significant kinematic variations with spatial position (Cantalupo, et al. 2005; Kollmeier, et al. 2010). As we discussed in Paper I (Martin, et al. 2013), we have therefore developed and extensively tested an adaptive smoothing algorithm (3DASL) designed to highlight extended line emission while not artificially smoothing more compact emission line and continuum sources. We repeat some of the discussion in Paper I here, and elaborate on simulations that are specifically designed to be representative of the LAB2 data cube. The goal of the following analysis is to search for extended emission regions associated with LAB2. This is a key step in deducing that LAB2 is centered on a nexus of filamentary, extended emission regions.

### 3.3 Smoothed Data Cubes: Extended Emission Kinematically Linked to LAB2

In Figure 6 we showed the CWI summed difference spectra obtained in several of the radial/azimuthal bins. In the first three significant emission is detected in an extended region beyond the blob (r>7 arcsec), and the line kinematics show some connection to those of the blob.

Based on NIR observations of [OIII] from the LAB2, the systematic velocity of LAB2 is near z=3.097 or Lyα redshifted to 4982Å. In some cases (e.g., panels a, b, e, g) double-peaked profiles may be present having the form predicted for optically thick Lyα (Neufeld 1990).

In Figure 7 we show images obtained for 4Å wide slices resulting from adaptive smoothing centered on the six velocity bins given in Figure 6. A description of this algorithm is given in Paper I and in the Appendix A.1. Extensions are detected outside the LAB2 region. Panels 3-6 shows that two galaxies with known redshifts in the field may be bridged by an extension. Comparison of the Figure 7 and Figure 2 (as well as Figure 4a and b) shows at least two common features: North-West (azimuth 30°-60°, radius 10-25 arcsec) and South-East extensions (azimuth 190°-240°, radius 10-25 arcsec).

We show in Figure 8 pseudo-slit spectral image plots obtained from cuts of the smoothed cube for a full set of slit position angles centered on LAB2. We have selected three position angles for detailed study in Figure 9 (displayed on the narrow-band image Figure 2). In each case extended emission is clearly detected. In the case shown in Figure 9(a), the emission is clearly double-peaked, and the peak separation increases approaching the blob. Double or triple peaked emission appears in all three slits of Figure 9. The central wavelength, peak separation, and peak intensity asymmetry varies somewhat with position. The morphology displayed in these spectral-image plots, double and multiple peaks with kinematic variations, is very similar to that predicted in simulations of Lyα emission from the cosmic web including realistic radiative transfer and velocity shear (Cantalupo, et al. 2005; Kollmeier, et al. 2010). Because the emission shows kinematic variations, but appears typically in bands less than 12Å wide, we developed a method to visualize the extended morphology in a single image, shown in Figure 10. These plots are generated to extract extended emission with common but possibly continuously varying kinematics. The intensity plot in panel (a) is generated as follows. At each azimuth (on a 7.5° grid), a spectral image plot like that in Figure 8 is generated (with a 5 arcsec slit width except for r<5 arcsec for which the slit width is reduced to 2.5 arcsec). For each position along the pseudo-slit, a moving window 12Å wide (20Å for r<5 arcsec) is permitted to slide over the full range of LAB2 velocities (4960-5000Å). The position with the maximum total emission within the window is selected. This is summed into the average azimuthal plot. In panel (a) the region with r<5 arcsec has a larger intensity scale (full-scale 50,000LU) to show LAB2 structure. Panel (a) shows that the extended emission is organized into azimuthal regions 30°-60° (defined as filament 1), 195°-240° (filament 2), and 330°-360° (filament 3). Panel (b) shows the mean wavelength calculated with a weighted sum in the sliding window. Values are shown only when the intensity in panel (a) exceeds 8000LU. Figure 10 shows that while the extended emission is complex and ubiquitous, it is principally organized into three azimuthal zones. Each of the three zones have, with some exceptions, kinematic spreads that are more restricted than that of the LAB2 and the extended regions taken as a whole. Figure 6, Figure 9, and Figure 10 suggest that the kinematics of the extended regions are correlated with the kinematics of the blob at similar position angles. In particular the south-east side of LAB2 is blueshifted, and the north-west redshifted. The extended emission displays a similar pattern. When the angular momentum is estimated (using the mass estimate discussed below), the vector directions for the blob and the extended emission are roughly parallel.

The extended Lyα emission appears diffuse and distributed in a filamentary pattern, and, in some cases, has a line-profile consistent with Lyα emission from a single layer of optically thick gas. We therefore denote these three extended, azimuthally and kinematically organized emission zones as filaments. We discuss in the next section and ultimately dismiss several other alternative explanations for this result.

**4. Alternate Explanations and Confirming Evidence**

We have investigated various explanations for these detections, which are at about 1-2% of sky continuum surface brightness. We investigated extensively the robustness of the smoothing algorithm, the detectability of extended filaments data cubes, and the impact of known compact sources. A full discussion is presented in Paper I and in the Appendix. We consider here sky subtraction errors, resolved source contributions, and radio-powered nebulae.

Residual sky subtraction errors have been minimized by the nod-and-shuffle technique, as discussed above. We estimate that residual errors should be <0.1%, or <200LU/Å. Sky subtraction and sky-line centroid errors would, to first order, be manifest as a uniform background change from one slice to the next. A Littrow ghost from the volume phase holographic grating was placed in a spectral region outside the band of interest (<4960Å).

The continuum objects in the field produce a small continuum background. We have already noted that subtracting these objects has no discernable effect on smoothed images and on the emission region spectra. This indicates that direct Lyα from the objects (if they are at the correct redshift) is not responsible for the signal. Lyα emitting galaxies exhibit Lyα halos (Steidel, et al. 2011) that summed, for the three galaxies between the two associated galaxies at z=3.091 and z=3.094 , for instance, could produce a smoothed diffuse level of at most ~100 LU/Å. Neither galaxy nor halo Lyα profiles are observed to be double-peaked. As we showed above, some or most of the apparent discrete clumps in Figure 7 are likely to be a result of the relatively low noise threshold for the smoothing algorithm (which show similar numbers of "clumps", e.g. Figure 23), but we consider the possibility below that they are significant.

Compact sources of narrow-band line emission have been surveyed in this field and a luminosity function calculated (Rauch, et al. 2008). Our sensitivity to compact sources is ~$1.3 \times 10^{-18}$ erg $cm^{-2}$ $s^{-1}$ (3-sigma). Integrating the continuum luminosity function and assuming that 20% of the sources have EW(Lyα)=20Å we find that we expect 0.6 sources in our field of view. Integrating to continuum magnitude $m_{AB}$<30 assuming a Schechter function slope of -1.6, we calculate that a total Lyα flux of $3.7 \times 10^{-18}$ erg $cm^{-2}$ $s^{-1}$ should be present in the field from these sources. This should be compared with the filament fluxes of (2-4) x $10^{-16}$ erg $cm^{-2}$ $s^{-1}$ and the total filament flux of ~$1.4 \times 10^{-15}$ erg $cm^{-2}$ $s^{-1}$ (see Table 1).

The faint population of Lyα emitters (Rauch, et al. 2008) would produce only 0.4 sources in the field of view and redshift range of this observation, on average. Their redshift distribution shows overdensities in 50Å bins as high as 4, so we could expected 1-2 such sources in the LAB2 overdensity, consistent with the calculation in the previous paragraph. With a typical source brightness of $10^4$ LU in 3 $arcsec^2$ such a source would contribute less than 100 LU/Å spread out over a 100 $arcsec^2$ filament area. A large population of very low luminosity star-forming galaxies that produce Lyα but are undetected in continuum or narrow-band images cannot be directly ruled out, but would have to have unreasonable volume densities and/or an even steeper faint end luminosity function than used above.

It is certainly possible that the detected (and fainter) sources are star forming galaxies that produce Lyα, and because their haloes are optically thick most of the Lyα leaks out far beyond the galaxies and traces the large-scale gas distribution. Much deeper and higher spatial resolution data and high resolution modelling are required to eliminate this possibility. However, the question of the radiation source can be separated from the morphology and kinematics of the extended, filamentary gas.

Some recent simulations (using smooth particle hydrodynamics) of Lyα emission predict a clumpy morphology that is consistent with our images (Kollmeier, et al. 2010). In particular, typically ~20 clumps are apparent in each 4Å slice (Figure 7), although many of these are produced by noise. A simular number appear in the most significant overdensities in the simulation over a similar field of view. Clump emission strengths are also roughly consistent for the QSO illuminated case.

Finally, (Steidel, et al. 2000) gave arguments against standard extended object explanations for Lyα blobs, such as extended radio source nebulae; those arguments stand for these much larger emission regions. The emission deficit at 4982Å has been attributed to absorption by a galactic superwind (Wilman, et al. 2005). Such a wind could in principle produce extended emission nebulosity if, for example, the outflow was striking inflowing IGM gas, although most of the energy would probably be deposited at x-ray temperatures. Furthermore, the morphology does not appear bipolar. While we cannot rule out an outflow as an explanation of some aspects of the emission morphology, the common kinematic distribution of the LAB2 absorption feature and the quiescent emission lines in the vicinity suggest that the LAB2 absorption feature is due to surrounding IGM gas. It may be possible with a detailed radiative transfer model to distinguish gas in the foreground and background, and thereby determine whether the gas is inflowing or outflowing.

**5. Discussion – What is the Nature of the Extended Emission?**

We have established that the extended emission is not an artifact of the data analysis or observational errors. It is unlikely that the emission is the smeared out result of many faint, compact sources, although definitive observations remain to be made. We therefore consider two distinct alternatives, that the emission is tracing gas from outflows produced by the galaxies in LAB2, or that gas that is inflowing from the cosmic web. We discuss in turn the morphology, energetics, and kinematics of the gas in the context of these two models. We note that it may be the case that inflowing and outflowing gas is present, and there are some hints that this is the case.

### 5.1 Morphology and Kinematics – Inflows vs. Outflows

While the distribution of the extended Lyα emission is complex, the concentration of emission into roughly three azimuthal confined extensions is suggestive of filaments of the cosmic web. Both the spatial morphology and line profiles resemble simulations of Lyα emission from the cosmic web (Furlanetto, et al. 2003; Cantalupo, et al. 2005; Faucher-Giguère, et al. 2010; Kollmeier, et al. 2010) and simulations of cold accretion flows (Dekel et al. 2009). As we highlighted in Figure 9(a), filament 2 displays a double peak centroid similar to the systemic velocity and the central (absorption) wavelength of the blob. The peak separation increases approaching the blob, consistent with an increasing optical depth, density, and/or velocity dispersion as could be expected in an inflow. In most cases, the line profiles and peak separations are typical for optically thick gas at temperatures of $2 \times 10^4$K predicted for the diffuse IGM.

The relative strength of the two peaks in the double peaked profile can be used in principle to determine whether the gas is inflowing or outflowing (Verhamme, Schaerer, & Maselli 2006) but as we illustrated with the simulations at the present signal-to-noise ratio while we can delineate filaments we cannot strongly constrain the line ratio on small scales. The large-scale line ratio appears to be near unity in the cases highlighted in Figure 9, which rules out large velocity offsets. With typical outflow wind velocity gradients, the double-peaked profile dissapears entirely (Verhamme, et al. 2006). Since the double-peaked profile is clearly detected in the data, this suggests that the velocity offsets and shears are incompatible with outflow kinematics.

Galactic or AGN outflows could produce shocked Lyα emission from entrained wind material but would likely yield broader line profiles and peak separations.  It would be surprising if the position angles of outflow emission was perpendicular to the blob spin axis (see §6.3). We also note that the significant amount of gas at the systemic velocity (4981.7Å) could be responsible for the absorption feature observed in LAB2. The emission deficit at 4982Å had been attributed to a superwind (Wilman, et al. 2005), but we would expect a feature blueshifted from systemic by 500-1000 km/s in that case. Such a wind could in principle produce extended emission nebulosity if, for example, the outflow was striking inflowing IGM gas, although most of the energy would probably be deposited at x-ray temperatures. Furthermore, the morphology does not appear bipolar. In the case of filament 2, it is hard to imagine scenarios in which the velocity dispersion decreased moving outward, although the density is likely to decrease. While we cannot rule out an outflow as an explanation of some aspects of the emission morphology, the common kinematic distribution of the LAB2 absorption feature and the quiescent emission lines in the vicinity suggest that the LAB2 absorption feature is due to surrounding IGM gas.

We note in fact that there is a bright blob near LAB2 at azimuth 130°, radius10-15 arcsec, and another at 20-25 arcsec. As we show below this is in fact orthogonal to the spin axis of the system. In Figure 11 we show that these small neighbor blobs appear at 4953Å, or $\Delta v$=-1700 km/s. This outflow velocity is large for a star forming galaxy but not unprecidented for a QSO. The proximity of the blobs to LAB2 could be a coincidence, but examination of the narrow-band image even suggests a shell-like appearance. We note however that there is a concentration of galaxies at this velocity so this could be a coincidence.

**5.2 What energy source powers the extended emission?**

Table 1 summarizes typical fluxes and luminosities obtained in azimuthal zones using the unsmoothed data cube. The typical flux is in the range of 600-1300 LU/Å, or about 5000-10,000 LU in 8Å bins. This corresponds to a line flux of $\sim 1 \times 10^{-18}$ erg cm$^{-2}$ s$^{-1}$ arcsec$^{-2}$. These are about 10-20 times the predicted level for Ly$\alpha$ fluorescence due only to excitation by the UV metagalactic background (Gould & Weinberg 1996b; Cantalupo, et al. 2005; Kollmeier, et al. 2010). Two explanations for the increased emission exist: boosted fluorescence and gravitational cooling. The radiation field of a nearby QSO can significantly boost Ly$\alpha$ fluorescence. LAB2 hosts an obscured AGN[25] with $L_X \sim 10^{44}$ erg s$^{-1}$. If the AGN is unobscured in the directions of the emitting filaments and has a typical spectrum, it could produce a boost factor $b$>100 over the full projected distance of ~200 kpc observed by CWI. Alternately, QSO SDSS J221736.54+001622.6 at z=3.084 is 3 arcmin to the north, and would produce an estimated boost of *b*~10 at LAB2. In this case it is likely that most of the emission is "reflected" rather than transmitted, and the double peak profile reflects the velocity dispersion of the illuminated gas, with $\sigma \sim c/4(\Delta\lambda_{1/2}/\lambda_{\alpha}) \sim 50$ km/s for $\Delta\lambda_{1/2} \sim 3.5$Å.

We used the nebular emission line code *Cloudy* (Ferland 1996) and a simple geometric toy model to derive rough, order of magnitude physical parameters in the Lyα emission filaments. For both models we assumed constant density, metallicity of 10% solar, unit filling factor, and an emitting region depth of $1.5 \times 10^{23}$ cm. For each model we varied the density, and for the collisionally ionized model we also varied the fixed temperature, to obtain the average observed Lyα flux of $1 \times 10^{-18}$ erg cm$^{-2}$ s$^{-1}$ arcsec$^{-2}$.

For the photo-ionized, Lyα fluorescence model, we used a standard metagalactic background model (Haardt and Madau, 2012) at redshift z=3 with a boost factor of 14. The flux saturates at a density of log $n_H$=-1.8, where the gas becomes self-shielding. This corresponds to total column density of log N(H)=21.4, and a (highly-uncertain) HI column density of log N(HI)=20.4. Higher column densities produce the same flux, since when the Lyman continuum is optically thick the gas is merely "reflecting" ~2/3 of the boosted ionizing radiation field, and therefor our results are lower limits to the column densities. The average electron temperature is $1.7 \times 10^4$ K. With a total filament area of 620 arcsec$^2$ (taking from Figure 10 the annulus 10-25 arcsec and a total azimuthal span of 135° from filament 1 (315°-360°), filament 2 (30°-60°), and filament 3 (180°-240°)) this results in a baryonic mass of at least $10^{11} M_{\odot}$.

Gravitational cooling radiation can produce the observed Lyα radiation, if the halo mass is high enough, because the majority of the baryonic gravitational binding energy is predicted to be dissipated as Lyα radiation in cold accretion flows (Fardal, et al. 2001; Birnboim & Dekel 2003; Furlanetto, et al. 2005; Kereš, et al. 2005; Faucher-Giguère, et al. 2010). A halo mass of $5 \times 10^{12} M_{\odot}$ yields an average flux of ~10,000LU ($10^{-18}$ erg cm$^{-2}$ s$^{-1}$ arcsec$^{-2}$) over a virial radius of 24 arcsec (Haiman, Spaans, & Quataert 2000), comparable to the average flux in the extended filaments. (Some recent simulations predict lower fluxes, depending on physical assumptions and halo mass, e.g., (Faucher-Giguère, et al. 2010). In this case the line profile would be that expected from a buried source of Lyα. Line center optical thickness and HI column density scale as $\tau_0 \sim 10^5 (\Delta\lambda_{1/2})^3$ and $N(HI) \sim 3 \times 10^{17} (\Delta\lambda_{1/2})^3$ cm$^{-2}$ for a standard diffuse IGM temperature of $T = 2 \times 10^4$ K (Neufeld 1990; Cantalupo, et al. 2005). With $\Delta\lambda_{1/2} \sim 3.5$Å, $N(HI) \sim 10^{20}$ cm$^{-2}$ (or a factor of 1.4 higher if the temperature is $10^4$K). This is consistent with the largely neutral gas and high Lyα optical depths suggested in the cooling models and cold accretion simulations (Dekel, et al. 2009). This halo mass would imply a circular velocity $v_c \sim 400$ km/s and a virial radius of 140 kpc or 17 arcsec.

For the cooling model we assumed coronal equilibrium and kept both density and temperature fixed. This of course is an approximation since the gas is assumed to be cooling and thus almost certainly displays a range of temperatures, as well as violating equilibrium conditions. The case that matches the observed flux and derived HI column density is log $n_H$=-2.25, log T=4.2, giving log N(H)=21.9 and log N(HI)=20.2. The baryonic mass is then $4 \times 10^{11}$ $M_\odot$ in Lyα emitting filaments which implies a dark matter halo mass of at least $2 \times 10^{12} M_\odot$, consistent with the mass derived above given uncertainties in the emission model and the possibility that a substantial fraction of baryons are hot (or cold) and invisible. This halo mass would imply a circular velocity $v_c \sim 300$ km/s not inconsistent with the LAB2 kinematics, and a virial radius of 100 kpc or 13 arcsec.

**5.3 Kinematics and Angular Momentum**

LAB2 falls in a saddle between two sub-concentrations, one toward the north-east and one to the south-west (Steidel, et al. 1998). LAB2 is possibly assembling from IGM gas flowing in from both kinematic groups. The gas, if collapsing in a single dark matter halo, has significant angular momentum ($L_{fil} = 5 \times 10^{15} M_\odot$ km/s kpc, using the mass from the cooling model) in the NE direction. This is consistent with a halo mass of $2 \times 10^{12} M_\odot$ which for a canonical spin parameter λ=0.05 would have an total (baryonic) angular momentum of $5 \times 10^{16} \left(8 \times 10^{15}\right) M_\odot$ km/s kpc. LAB2 has gas angular momentum with a similar orientation ($L_{LAB2} = 2 \times 10^{14} M_\odot$ km/s kpc). This demonstrates in yet another way that the extended gas is kinematically related to LAB2.

We can also estimate the mass flux into a virial radius, by assuming all velocities within 500 km/s of systemic represent infall (i.e., redshifted gas is in the foreground and blueshifted in the background). The result is 200 $M_{\odot} yr^{-1}$ including LAB2, and one third of this excluding the blob. This is also consistent with the total baryonic mass inflowing within a local Hubble time of $10^9$ years.

A detailed kinematic model including radiative transfer will be presented in Paper III (Martin et al., in prep) with the goal of better constraining the geometry, radiative transfer and kinematics of the gas in this system.

**6. Summary**

1. We have detected extended emission around LAB2 which is kinematically associated with the LAB2 line profile.

2. The extended emission appears in the unsmoothed data cube and in adaptively smoothed image slices and spectral-images.

3. We have demonstrated with simulations that the smoothing algorithm is robust and provides a reasonable reproduction of extended emission with complex kinematic spatial and spectral profiles expected for emission from the cosmic web, in the presence of compact sources. The method does not artificial smear compact sources to produce extended emission.

4. The emission is detected with modest signal-to-noise ratio, and is very complex, but seems to be organized largely into three azimuthally confined extensions, which we call filaments, at azimuth 45°, 220°, and 330°.

5. The emission flux is 10-20 times brighter than that expected for the average metagalactic background powered Lyman alpha fluorescence. The emission could be produced by boosted fluorescence, possibly from an obscured QSO, or gravitational cooling radiation. In both cases the baryonic mass is roughly $4 \times 10^{11} M_{\odot}$ and the dark halo mass is at least $2 \times 10^{12} M_{\odot}$.

6. The LAB2 absorption is produced by gas in the vicinity at the systemic velocity of 4981.7Å. There is some evidence for inflow along the 30° and 210° azimuth. The spectral mophology appears inconsistent with an outflow.

7. An outflow may have been detected along the 135° azimuth, with Δv=-1700 km/s.

8. LAB2 and the surrounding gas have significant, but typical angular momentum for a halo of the predicted mass. The angular momentum of LAB2 is aligned with the angular momentum of the surrounding gas, strengthening the implied connection and the case for inflow.

## Appendix

### A.1 3D Adaptive Smoothing Algorithm

The slice images displayed in Figure 7 are produced from this difference flux cube by a 3D adaptive smoothing in lambda (3DASL) algorithm. The algorithm works as follows. The (0.292" × 0.292" × 1Å spaxel) data cubes are first smoothed spectrally by 2Å, then smoothed spatially, by successively larger boxcar smoothing kernels ranging from 5 pixels (1.4 arcsec) to 70 pixels (20 arcsec) in 2 pixel increments. The effective signal-to-noise threshold is correspondingly reduced by a factor equal to the one pixel noise divided by the 1D smoothing width (e.g., a 1-sigma 1-pixel noise of 9000LU becomes 1800LU for a 5 pixel smoothing box, and 180LU for a 50 pixel box). Whenever a spaxel exceeds the 2.5-sigma noise threshold (which of course depends on the kernel) it is placed in the final cube (detected). The flux in the smoothed spaxel is subtracted from the unsmoothed cube pixel prior to the next spatial smoothing cycle. Any residual flux in the spaxel can still contribute to the smoothed slice images but is not available for further "detection". The unsmoothed difference cube is then smoothed in wavelength by a 4Å kernel, and the spatial smoothing algorithm is repeated. This loop repeats once more with a spectral smoothing of 8Å, or a total of three times. The 2.5-sigma noise threshold was derived directly from the difference cube, and is consistent with the predicted Poisson noise. We discuss below variations in the algorithm and signal-to-noise thresholds.

We show in Figure 12 images that illustrate the raw data and noise in comparison to the adaptively smoothed image slices, for the same wavelengths as shown in Figure 7. In Figure 13 we show that pure noise produces a very different smoothed image and emission morphology. The noise cube was generated using an algorithm which simulates an input image and feeds it into the data reduction pipeline (DRP). The input image combines an appropriately distorted and sliced sky spectrum with simulated source filaments and compact sources, adding detector read noise and Poisson noise to detected photons. The input image is then run through the DRP. The data for this paper were obtained over three runs and a total of seven nights, in three different position angles, slightly different wavelength scales, and at multiple dither positions. The simulation images are produced with the necessary position angles and dithers, and then reconstructed into rectified data cubes, and finally mosaiced into a single coadded cube, using the same DRP software that is used on data. The principal effect is the digitization of data into 24 slices, which are later magnified by a factor of 9 to match the plate scale in the direction along the slices, and thus producing correlated errors. A second effect is the non-linear warping required to map curved slice images onto a rectilinear cube. This insures that the noise properties and any features introduced by the DRP are fully realized in the simulated cubes. In the case of Figure 13, no sources were added and the image consists of sky only. The appearance of the noise in regions without emission is quite similar in character to that of the actual data cubes.

To further amplify this point, we show histograms of the distribution of voxel intensity values in the unsmoothed, noise, and smoothed difference data cubes. In Figure 14(a-b), we show the historgram of the unsmoothed cube compared to that of the simulated noise cube. There is an excellent correspondance between the two, demonstrating that there is no significant source of systematic error present in the data beyond Poisson noise. The figure shows a single Gaussian fit to the data cube as well. A single Gaussian is a good fit out to several standard deviations, then falls below the observed pixel distribution. As the figure illustrates, this deviation is explained by the multiple regions of varying signal-to-noise ratio because of the mosaic and the variations in instrument throughput. These effects are well reproduced in the noise cube. In Figure 14(c-d) we show a comparison of the raw and 3DASL cube data histograms. The smoothed data is as expected positive definite and displays a tighter distribution.

An important point is that sky subtraction could yield some significant negative residuals if emission associated with the LAB2 large-scale structure is present in the “sky” region. The “sky” region is ~2 arcmin away, corresponding to ~1 Mpc physical and ~4 Mpc comoving from LAB2. Significant structure could remain at or near the LAB2 velocity at this distance. We choose a region with no obvious features in the narrow-band image, and we dithered the “sky” region extensive to average over smaller features. This unfortunately represents a fundamental limitation to this approach. We have checked using simulated emission cubes that this technique is not predicted to produce large subtraction residuals, assuming the structures in the "sky" region are of typical IGM brightness (e.g., consisting only of Lyman alpha fluorescence from the ambient radiation field). To check this we simply multiplied the difference cube by -1 and ran the identical 3DASL algorithm on the inverted image. The result is shown in Figure 17. There is no significant emission detected beyond that expected from noise fluctuations.

It is important to note that because of the modest signal-to-noise ratio of the emission there will not be a detailed one-to-one correspondance of the smoothed slice images and the underlying data, particularly on small scales. We used extensive simulations of filaments in Paper I to illustrate this point, and provide further examples representative of the present case below. As we show below with representative simulations for the LAB2 field, the algorithm tends to produce less extended and more clumpy emission than is underlying, again due to modest signal-to-noise. However as we will demonstrate there is a good correspondance on the scales of interest for this investigation.

Data in slices outside the velocity window occupied by LAB2 and large scale structure in the vincinity could in principal be used to provide a baseline with which to compare the detected emission. Unfortunately only the central 80Å are available because of the staggered image slicer design, which offsets each slice alternately +/-35Å. The smoothing algorithm requires a buffer of another +/-5Å, so only 4940-5010Å can be analyzed safely. We show in Figure 15 the complete set of image slices in this range. A particularly strong complex is present in the 4950-4956Å range. In Figure 16 we show a histogram of galaxy redshifts in the SSA22 area, based on publically cataloged galaxies. There is significant structure in the 4950-5005Å range, and the emission detected in Figure 15(b-c) may be associated with a structure that appears around 4954Å in Figure 16.

**A.2 Compact sources--impact on results.**

A large number of faint compact sources could in principle be converted by the smoothing algorithm into apparent extended emission in the smoothed images. Such sources if unsubtracted could also add spurious emission to the unsmoothed data cube and spectra derived from regions. This could be particularly worrisome for Lyα emission sources at the same redshift as LAB2. We have therefore carefully examined the impact of compact sources on the results. We used a continuum image to determine the location and approximate brightness of all compact continuum sources in the field of view of our source field. We determined the point spread function (PSF) from prior observations. In this PSF 80% of the light is confined to a region of 7.4 $arcsec^2$. We used a source removal algorithm that uses this PSF and neighboring sky, slice by slice, to remove each source and replace it with an estimate of local sky (which could include local extended emission). There are a total of ~40 sources detected in the continuum image of the source field to $V<25.5$. Figure 18 shows the compact source distribution and magnitudes superimposed on the narrow-band image, along with the boxes used for source definition. There are no significant differences in the resulting smoothed images or region spectra with or without source subtraction. Figure 19 shows an image comparison with and without sources subtracted. All images and spectra are given with sources subtracted.

**A.3 Modification to smoothing algorithm.**

The smoothing algorithm smooths in space (between 5 and 70 pixels or 1.5-20 arcsec) and in wavelength (2Å, 4Å, and 8Å). Adding an additional 16Å step makes little difference. In many cases a given pixel will be detected in multiple instances of a given spatial smoothing with different wavelength smoothing. The algorithm must choose which of these multiple detections to select. Also, the detected emission is removed from the image prior to successive smoothing cycles, so this choice and the order of considering the four wavelength-smoothed images impacts successive iterations. We therefore investigated the impact of different detection priority and wavelength smoothing order. We found that changing the detection priority makes only very small differences in the smoothed image. Reversing the wavelength smoothing order makes a slight difference and can produce a slightly less noisy appearing image, but reduces the contrast of the minimum between two emission peaks and is therefor less representative.

**A.4 Modifications to the signal-to-noise threshold.**

For a pixel to be detected it must exceed the signal-to-noise threshold. The choice of the signal-to-noise threshold has a moderate effect on the resulting smoothed images. Our extended emission detections are modest in signal-to-noise ratio, so we have investigated the impact of this threshold on the resulting smoothed images. In Figure 20 we show how the results in three slices varies with S/N threshold from S/N>3.0 to S/N>2.0. Later we show results with simulated filaments. The general trend in the lower S/N cases is that more compact regions are detected first and reduce the amplitude of more extended emission. Based on our study of simulated scenarios (below) we conclude that a conservative compromise is a threshold of S/N>2.5.

**A.5 Detection in multiple wavelength bins/slices.**

Examination of Figure 21, which shows the full set of 1Å 3DASL images for the band around LAB2, shows that most extended features are detected in multiple slices, and that many velocity features are detected in multiple regions. Filament 3 (see §4 for definitions) is detected in at least 4 wavelength bins (4972-4980Å), filament 1 is detected in 8 bins (4977-4980Å, 4984-4987Å), and filament 2 is detected in 7 bins (4966-4972Å). While the 3DASL algorithm does smooth over wavelength as well as angle, almost all noise features appear at most in 2 wavelength bins, as we illustrate in Figure 22.

**A.6 Simple filament simulations.**

We have performed simulations to investigate the behavior of the algorithm with and without compact sources, for difference algorithm variations and signal-to-noise thresholds. The first case we studied was a single filament of intensity and line profile similar to that of the detected emission. The filament has a total flux 15000LU in a double-peaked optically thick Lyα line profile with peak separation 3.5Å, velocity dispersion 60 km/s, filament width 12 arcsec, length 60 arcsec, position angle 40°, convolved with a 1.2 arcsec FWHM seeing disk. We include a set of compact sources in a “filamentary” distribution with fluxes comparable to those near LAB2. Figure 23 shows a set of these using different algorithms and signal-to-noise thresholds, with and without source subtraction. The source distribution is the same as the LAB2 field. We see that as the signal-to-noise threshold is reduced, more noise appears in the smoothed, thresholded image. The dominant character of this noise is a widely distributed set of compact noise spikes. Reducing the S/N threshold favors the compact noise spikes over the extended emission and is more sensitive to real compact sources. Compact sources have a modest effect on the smoothed image, slightly distorting the filaments and creating low-level extensions associated with bright neighboring objects. None of the displayed smoothed slices is significantly different in character from the ensemble. A threshold of 2.5 or 3.0 provides a good compromise between correctly characterizing the extended and the compact sources.

### A.7 Spectral Image Plots

In order to explore the spatial-spectral content of the data cubes, we generate pseudo-slit spectral image plots to show extended features that show single or multiple line profiles and possible kinematic gradients. These are generated by summing the adaptively smoothed data cube within a slit oriented at a particular position angle or in a free-form contour designed to follow an extended region. We show in Figure 24 examples of spectral image plots for the simple filament simulation introduced in the previous section. The spectral image plot of the smoothed cube shows good correspondance. The principal deficiency is that the line ratio in the double peak profile changes over the length of the slit, due entirely to noise.

To show the value of spectral image plots, we add to a simulation velocity gradients. The simulation displayed in Figure 25 has two filaments, both with velocity profiles that place them in multiple image slices. The pseudo-slits for the spectral image plots are oriented along the filaments. In the latter, we can clearly trace the varying velocity profile of the filaments.

### A.8 Representative filament simulations.

Finally we performed simulations that show a filament and source distributions schematically resembling that detected around LAB2. In Figure 26 and Figure 27 we show a four filament simulation in multiple wavelength slices. This illustrates that the algorithm can separate components centered on different velocities (wavelengths). In the simulation the southern filament is centered at 4974Å, while the other three are centered at 4982Å. In Figure 28 and Figure 29 we compare two cases--4 filaments and sources similar to that observed, and 4 short filaments that sum to form LAB2 and the same set of sources. The extended filaments are clearly detected in the first case, and not in the second case, in both image slices and spectral-image plots. These results are futher evidence that the smoothing algorithm correctly identifies extended filaments in the presence of compact sources and the central blob.

**Acknowledgements**

We thank Tom Tombrello and Shri Kulkarni for their support of CWI. We thank Marty Crabill, Steve Kaye and the staff of the Palomar Observatory for their constant support. Nicole Ligner participated in the observations. We are deeply grateful to Dean Joe Shepard, to the Caltech Counselling Office, and to the family of Daphne Chang for their strength and support. The anonymous referee provided excellent suggestions that significantly improved the paper. This work was supported by the National Science Foundation and the California Institute of Technology.

## Tables

Table 1. Derived Fluxes and Detection Significance for Azimuthal Regions Defined in Figure 2.

| Region | Azimuth | Radius | Area [Arcsec$^2$] | Band | Flux 1 | Flux 2 | Flux 3 | Flux 4 | Flux 5 | Luminosity [erg/s] | S/N |
|---|---|---|---|---|---|---|---|---|---|---|---|
| **LAB2** | **0-360°** | 0-7″ | 155 | 4950-5004Å | 1.1e-15 | 6.6e-18 | 1.2e-19 | 69120 | 1260 | 8.8e+43 | 23.3 |
| **Fil 1** | **30-60°** | 7-25″ | 150 | 4965-4981Å | 1.8e-16 | 1.1e-18 | 6.6e-20 | 11900 | 700 | 1.5e+43 | 6.3 |
| **Fil 1** | **30-60°** | 7-25″ | 150 | 4981-4996Å | 1.9e-16 | 1.2e-18 | 7.7e-20 | 13010 | 810 | 1.6e+43 | 6.8 |
| **Fil 2** | **195-240°** | 7-25″ | 223 | 4965-4981Å | 4.9e-16 | 2.1e-18 | 1.2e-19 | 22250 | 1310 | 4.1e+43 | 12.4 |
| **Fil 2** | **195-240°** | 7-25″ | 223 | 4981-4996Å | 2.1e-16 | 9.3e-19 | 5.8e-20 | 9770 | 610 | 1.8e+43 | 5.4 |
| **Fil 3** | **330-360°** | 7-25″ | 150 | 4965-4981Å | 2.0e-16 | 1.3e-18 | 7.6e-20 | 13680 | 800 | 1.7e+43 | 6.0 |
| **Fil 3** | **330-360°** | 7-25″ | 150 | 4981-4996Å | 1.8e-16 | 1.2e-18 | 7.2e-20 | 12300 | 760 | 1.5e+43 | 5.3 |
| **Filaments** | | 3 fil | 523 | 4665-4996Å | 1.4e-15 | 2.8e-18 | 8.9e-20 | 28400 | 920 | 1.2e+44 | |

Flux 1 = erg cm$^{-2}$ s$^{-1}$. Flux 2 = erg cm$^{-2}$ s$^{-1}$ arcsec$^{-2}$. Flux 3 = erg cm$^{-2}$ s$^{-1}$ arcsec$^{-2}$ Å$^{-1}$. Flux 4 = LU. Flux 5 = LU/Å. Errors in each unit can be derived by dividing flux by S/N.

## Figures

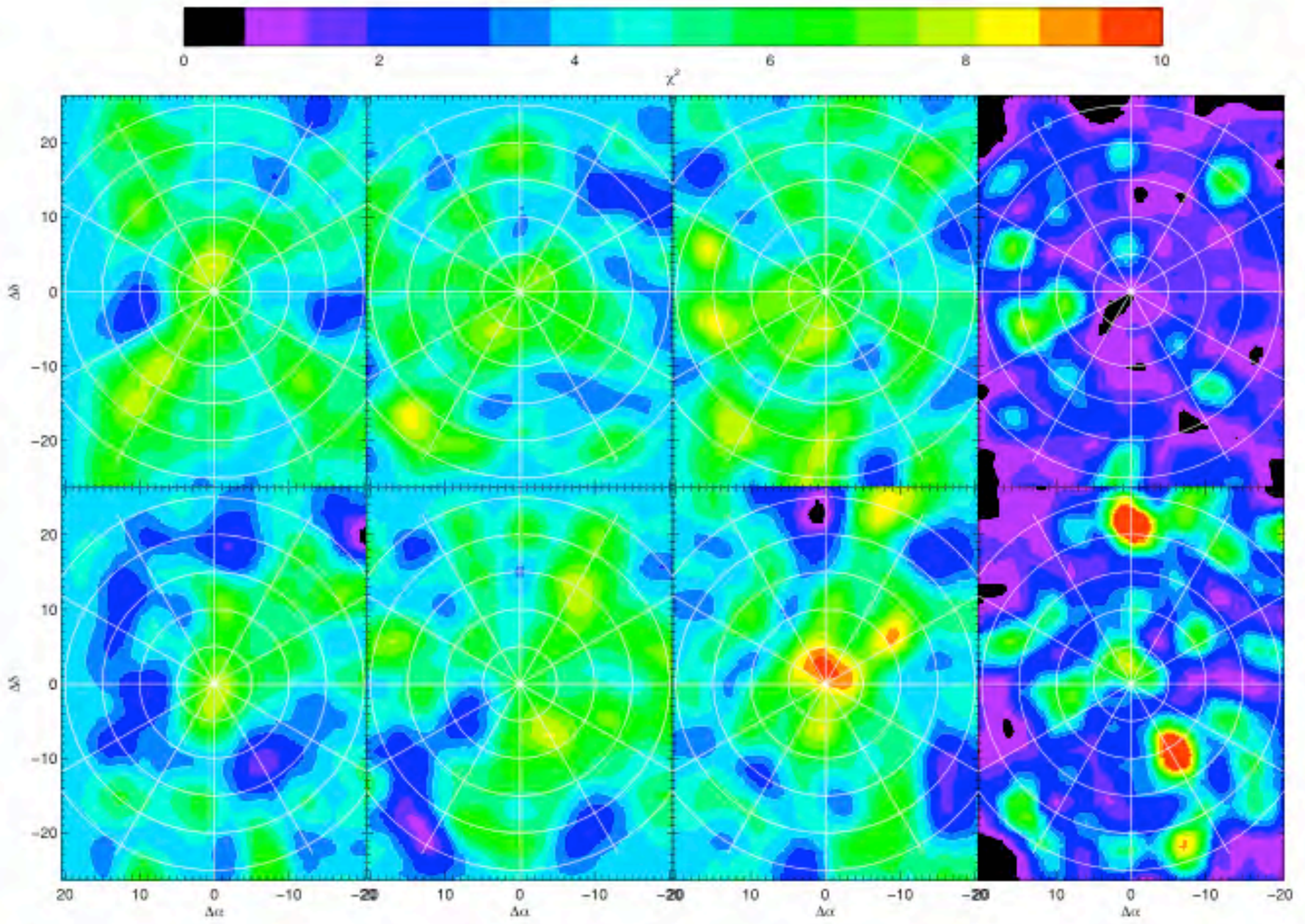

Figure 1. Statistical consistency of three CWI LAB2 runs. Each row gives a 4Å wide slice of the difference image, smoothed by 17×17 pixels (5 × 5 arcsec). Top row gives 4972Å, bottom 4986Å. From left to right, run 1 (3 hours source, 3 hours background, no nod-and-shuffle, single background region, position angle 0°), run 2 (6 hours source, 6 hours background, nod-and-shuffle, dithered background region, position angle 270°), run 3 (2 hours source, 2 hours background, nod-and-shuffle, dithered background region, position angle 180°). These panels are scaled so that the color scales runs from -4σ to +6σ. Right-hand panels give total chi-square for each slice in units given by the scale bar. The total chi-square is defined as $\chi^2(x,y)=\sum_{r=1}^{3}\left\{\frac{\left[I_r(x,y)-I_{tot}(x,y)\right]}{\sigma_r(x,y)}\right\}^2$, where $I_r(x,y)$ is the difference cube intensity for run $r$ (vs. position $x,y$), $I_r(x,y)$ is the coadd intensity, and $\sigma_r(x,y)$ is the rms error for that run. The distribution of chi-square over position is consistent with the chi-square probability distribution function. For roughly 100 independent smoothed patches (5 × 5 arcsec$^2$), 18 will exceed $\chi^2$=5 and 3 will exceed $\chi^2$=9. Thus chi-square distribution is within assumed variances for both slices. Azimuthal reference grid from Figures 1, 2, and 5 is also shown. The chi-square estimate does not include any error introduced from subtraction of sky spectrum that may incorporate some features present in the source spectrum.

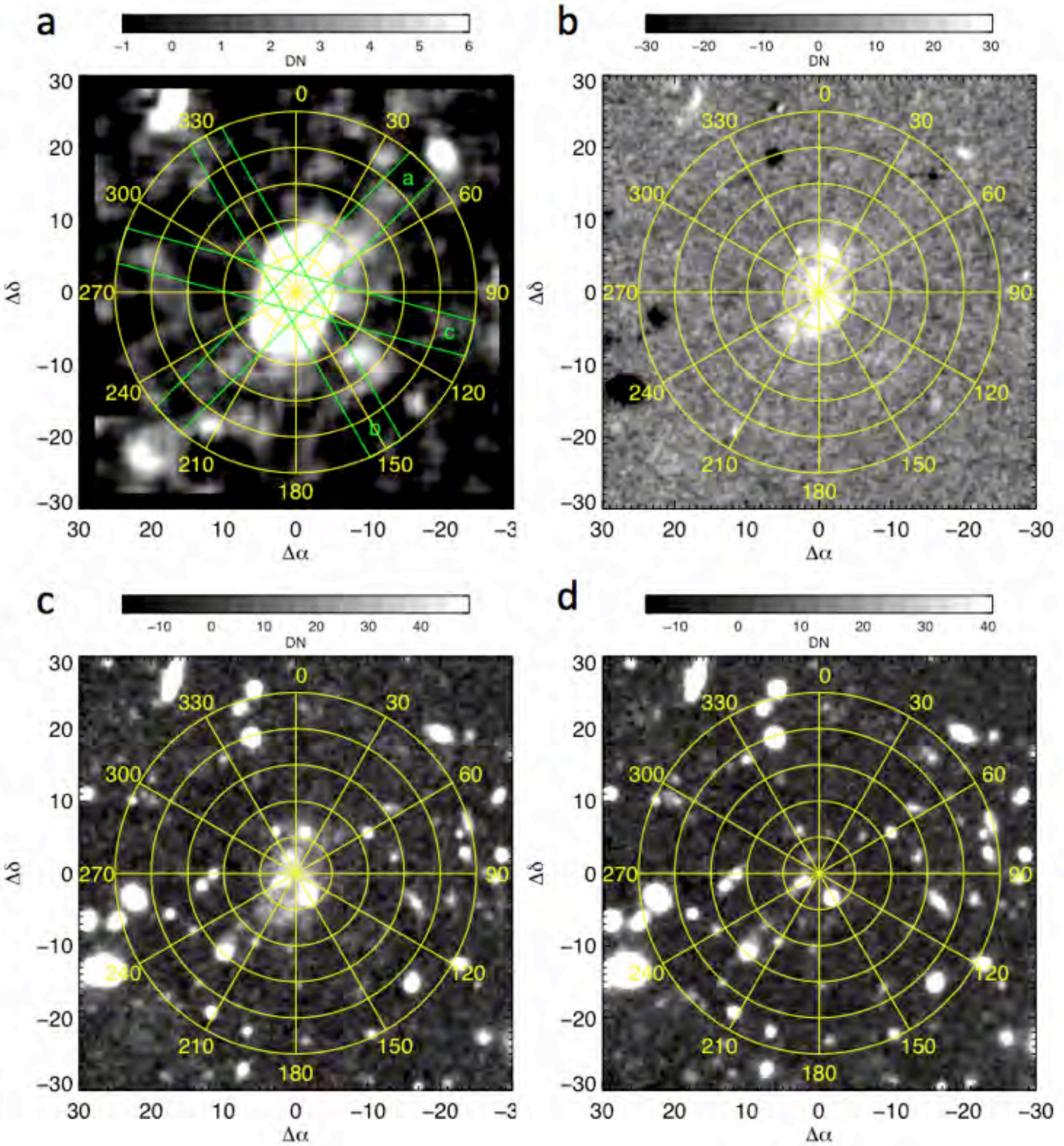

Figure 2. Narrow band imaging of LAB2. a. narrow-band, continuum-subtracted image of LAB2 (*NB479-BV, see text)* smoothed (5 arcsec kernel) and stretched in order to emphasize low-surface brightness extended features in the surrounding regions. The image was taken by Suburu Supreme-Cam (Hayashino, et al. 2004; Matsuda, et al. 2004), and consists of a total integration time of 7.2 hours in a 77Å wide filter centered on Lyα at z=3.094. We have superimposed a radial grid for reference and for extracting spectra from the CWI data cube. The grid consists of radial bins on 5 arcsec intervals, and azimuthal bins 30 degrees wide, centered on LAB2. In all images North is up and East is to the left. Pseudo-slits used to generate spectral images in Figure 9 shown in green. b. Unsmoothed *NB479-BV*. c. narrow-band image without continuum subtraction (*NV479*), unsmoothed. d. Continuum image *BV*, unsmoothed.

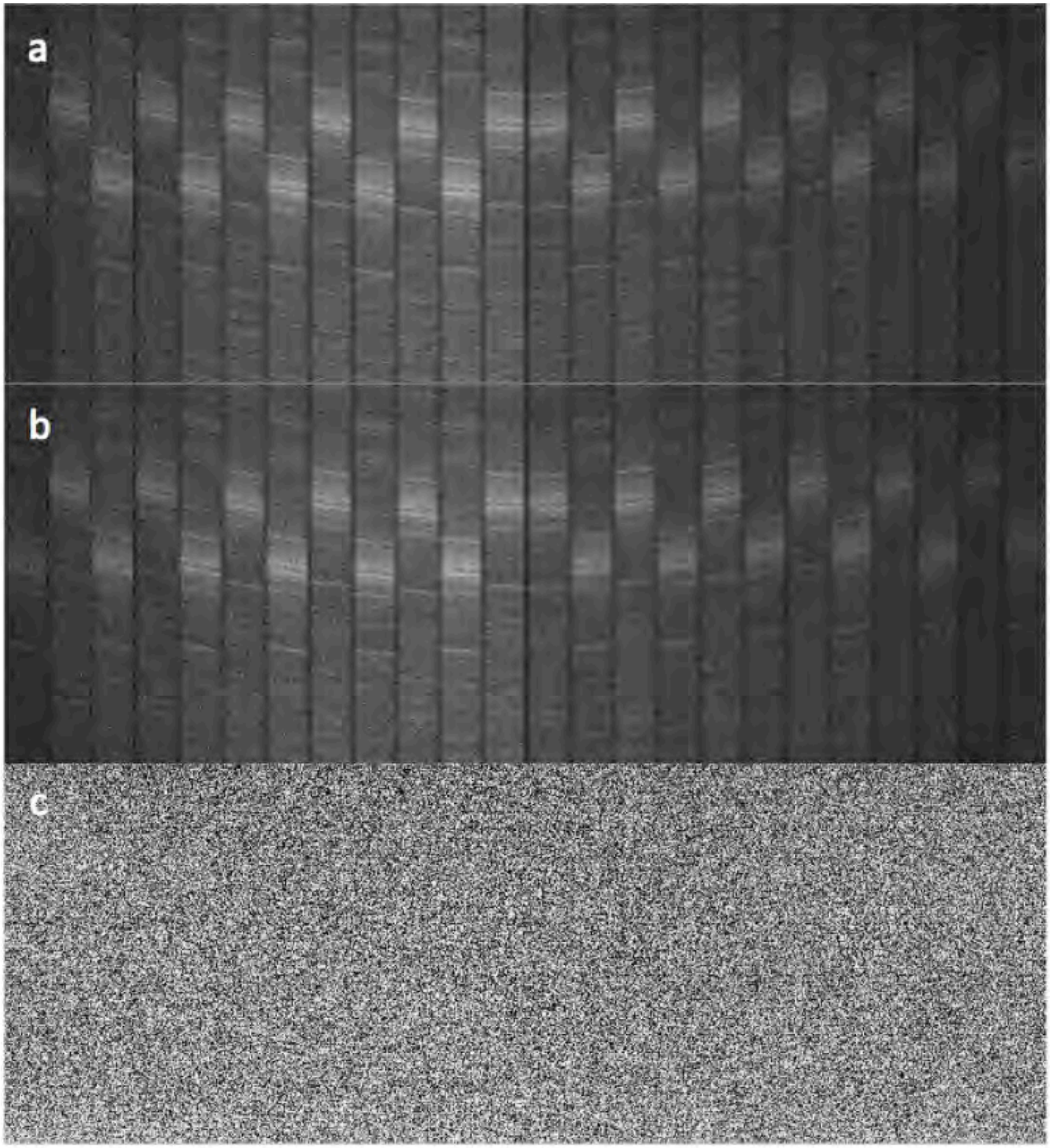

Figure 3. A single nod-and-shuffle frame. a. source region exposure, a total of 20 minutes, 10 exposures of 2 minutes each. b. background region exposure, a total of 20 minutes (see text). c. difference frame, intensity scale is +/-0.1*(scale of panel a/b). Frames have been biased subtracted, cosmic-ray cleaned, and slice extracted. No other processing has been performed.

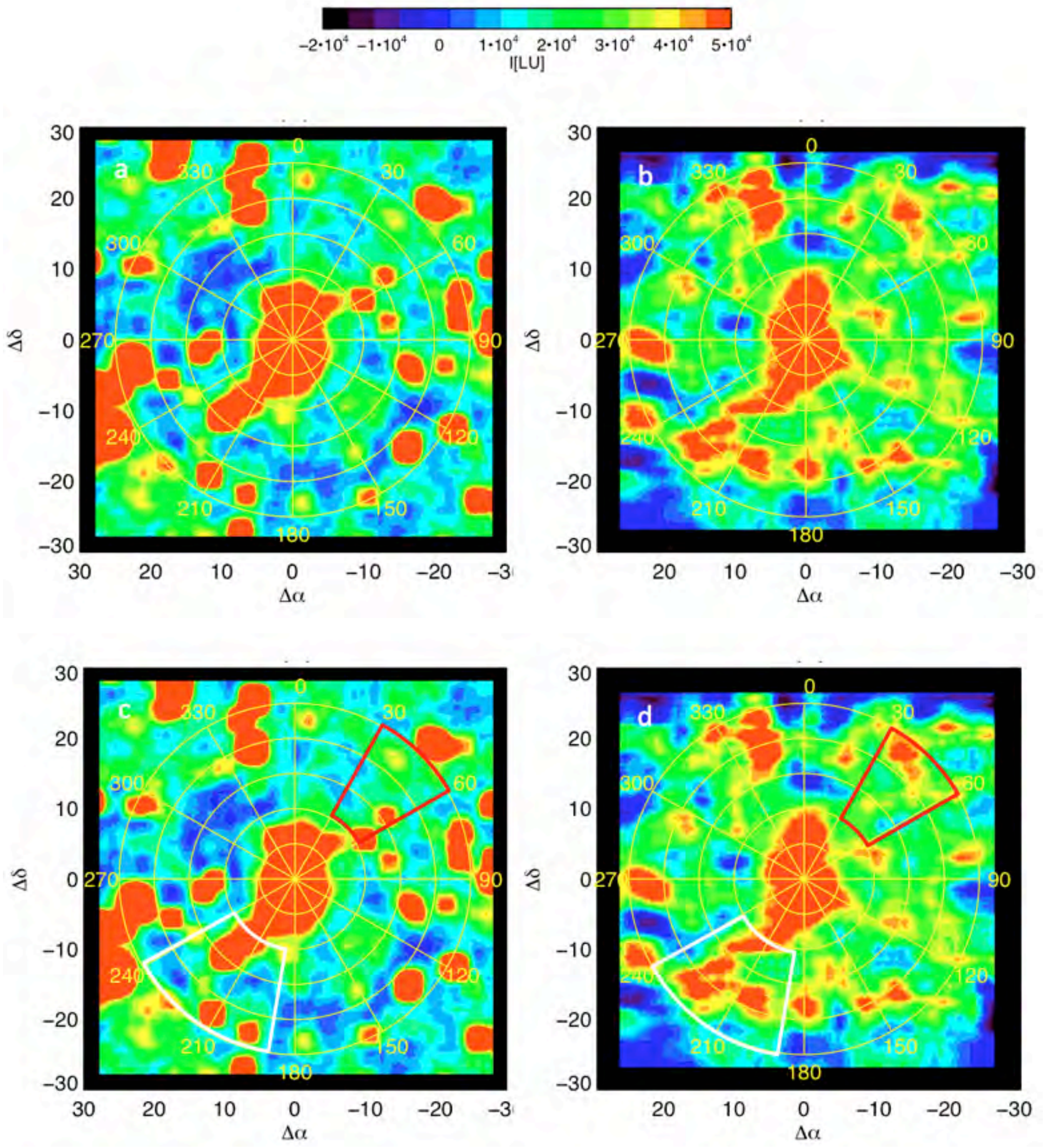

Figure 4. Comparison of narrow-band image (Subaru) and CWI stacked image. a. Narrow-band image (sky-subtracted but not continuum subtracted), now with difference color scale (shown), with smoothing 4 arcsec boxcar. b. CWI stacked, non-adaptively smoothed image with simple 4 arcsec boxcar smooth, same intensity color scale. Radial grid is superposed to aid in cross-identification of emission features. The narrow-band image has a nominal Poisson noise level of 2.5e-19 erg $cm^{-2}$ $s^{-1}$ $arcsec^{-2}$ (2500LU) for this smoothing. However, typical systematic

continuum-subtraction errors can be considerably larger, 7e-19 to 10e-19 erg $cm^{-2}$ $s^{-1}$ $arcsec^{-2}$ (7000-10000LU). The CWI stacked image has a similar Poisson error level (~7000-10000LU). Thus for both images typical noise amplitude will cause variations from blue to green color values on 5 arcsec scales, as is observed. Comparison of individual features detected in either image will not be one-to-one because of this noise level. c. and d. show same images with filament labels (1=red, 2=white).

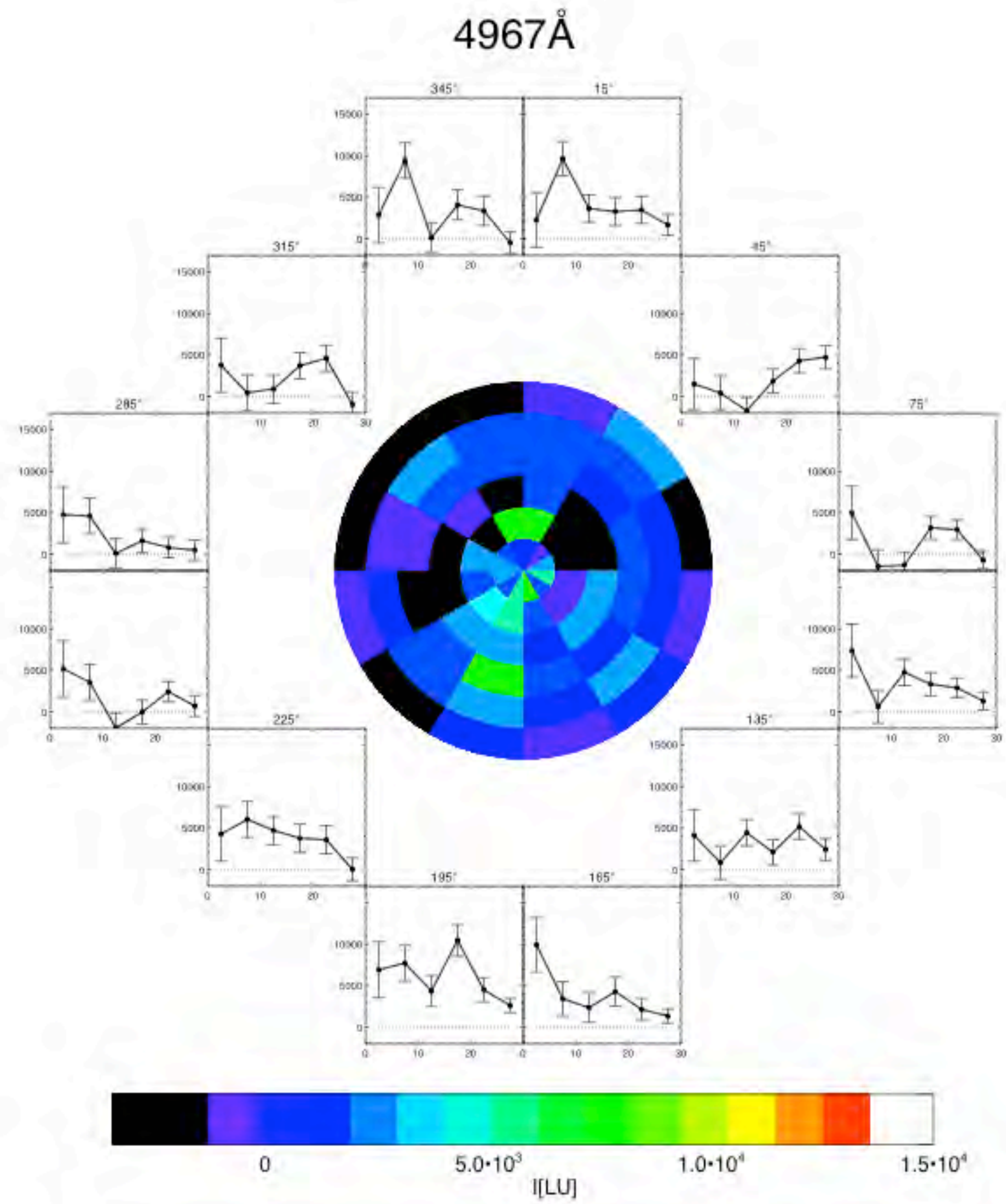

Figure 5. Unsmoothed 4Å data cube slices plotted in 30° azimuthal and 5 arcsec radial bins. Image shows intensity in each bin. Plots show radial profile in each azimuthal bin. a. wavelength bin centered at 4967Å. An extension is present in azimuth 195°. All intensities given in LU.

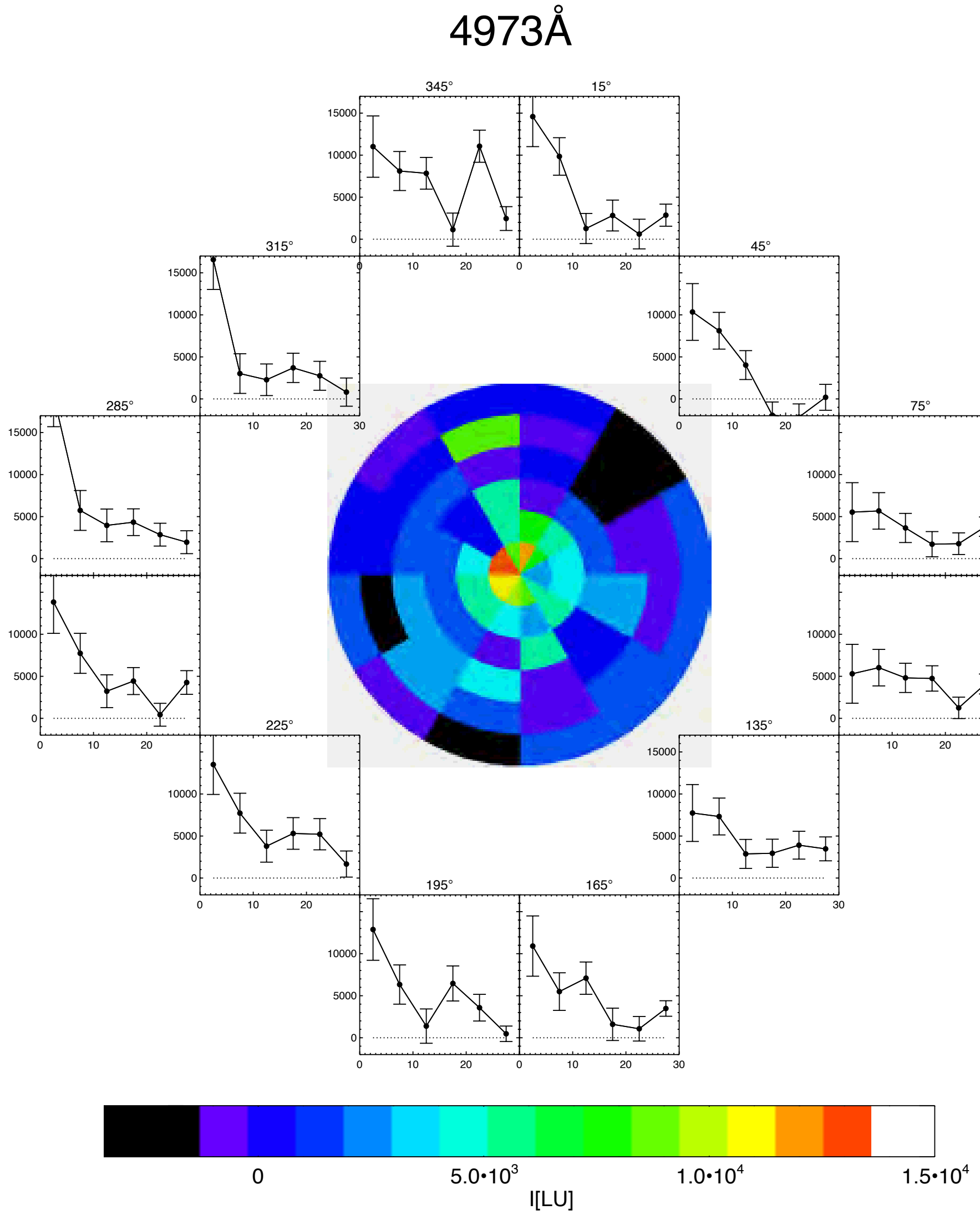

b. 4973Å wavelength bin. Prominent extensions present 180°-240° and 330°-360°.

4976Å

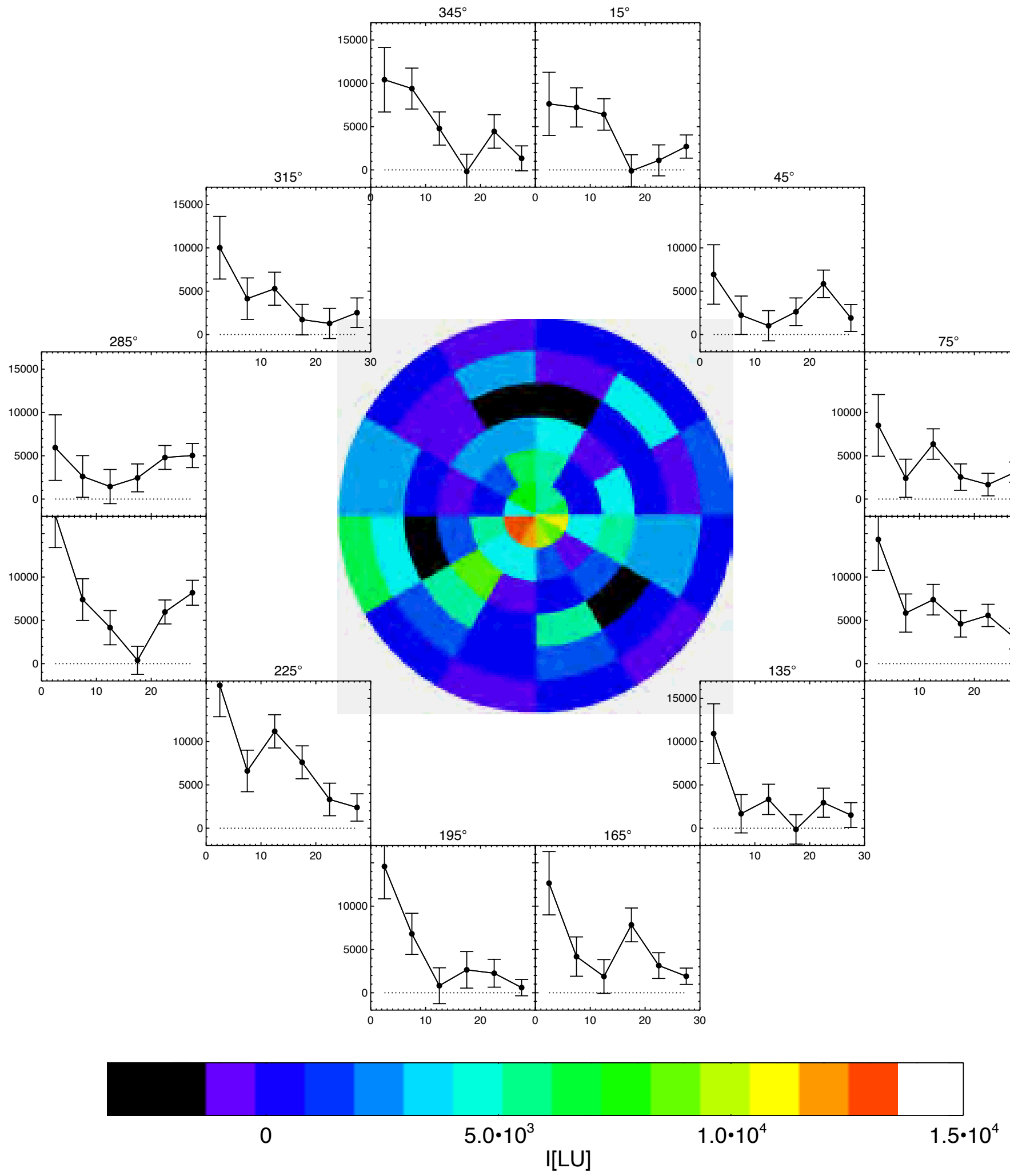

c. 4976Å wavelength bin. Prominent extensions present 210°-270°, 0°-60°, 105°, and 165°.

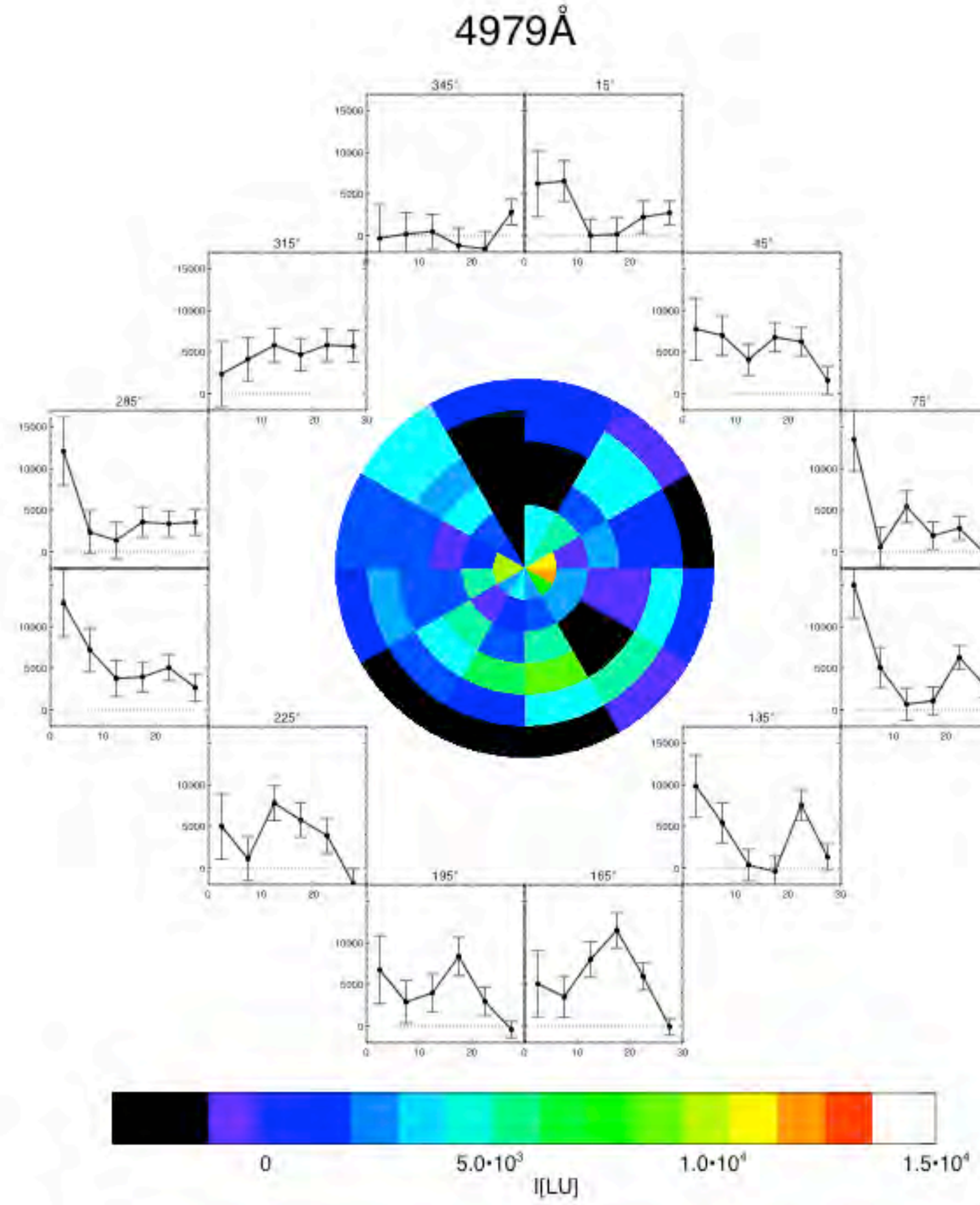

d. 4979Å wavelength bin. Prominent extensions present at 30-60°, 150°-240°, and 315°.

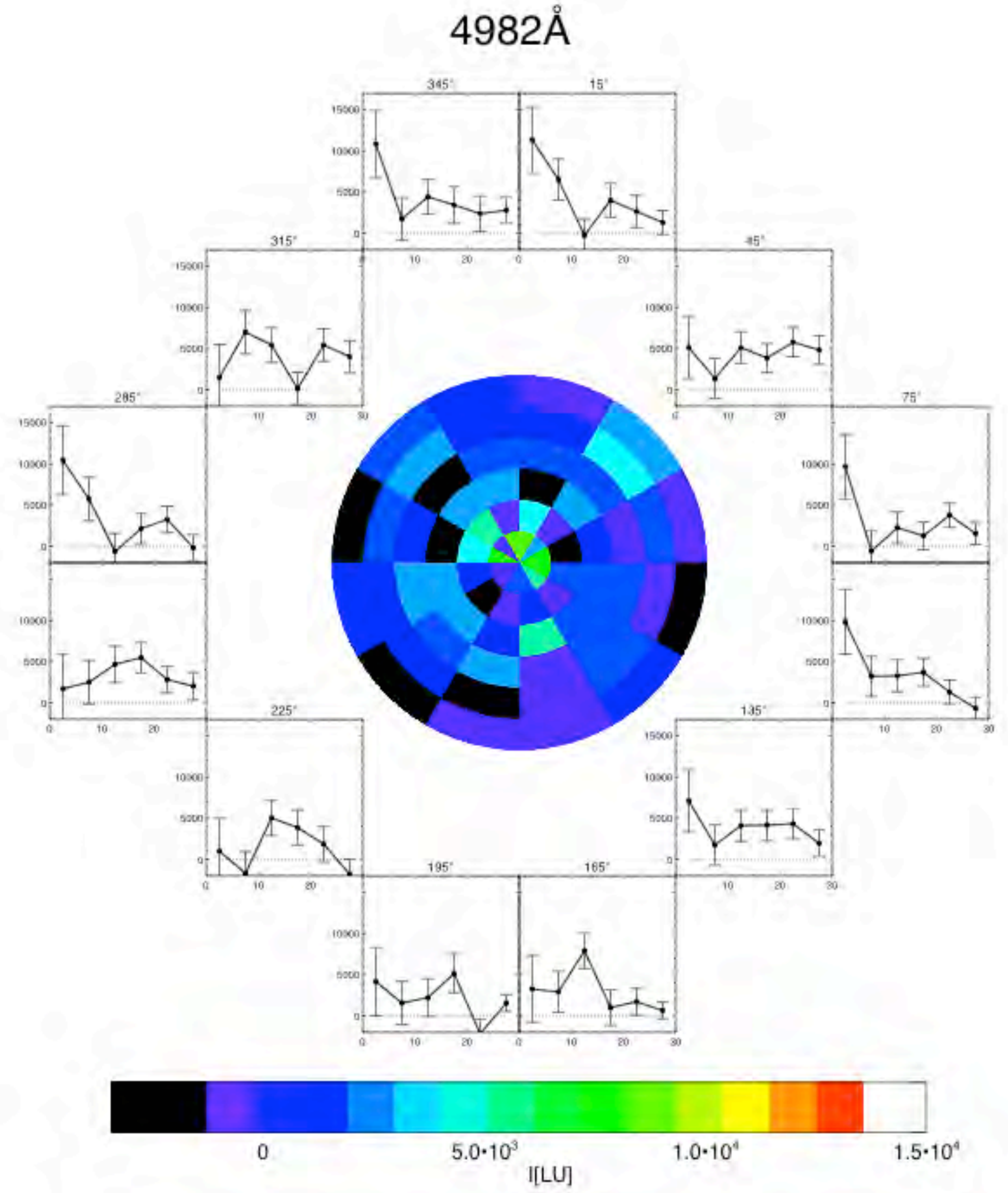

e. 4982Å wavelength bin.

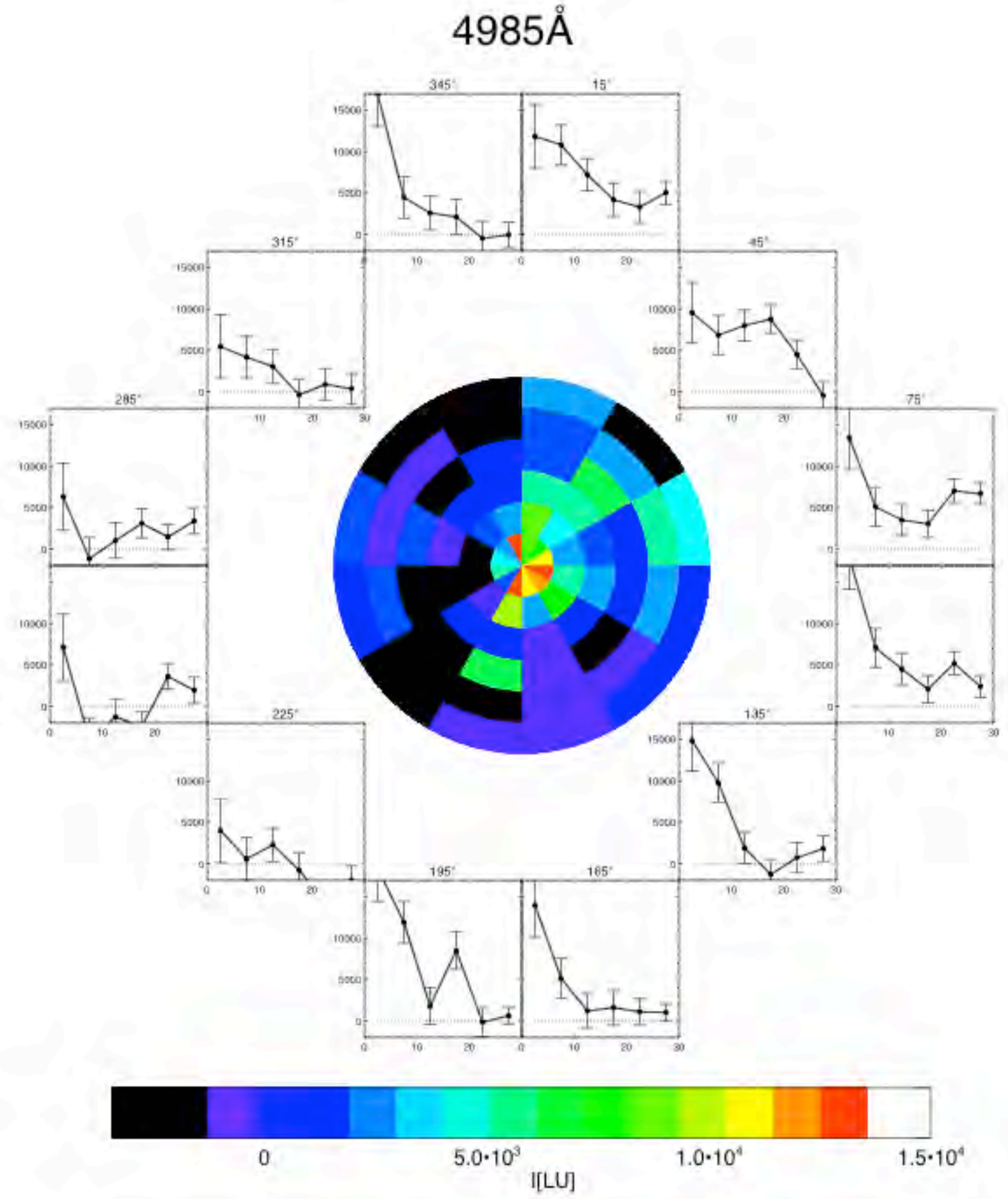

f. 4985Å wavelength bin. Prominent extensions 0°-90°, with some extension at 195°.

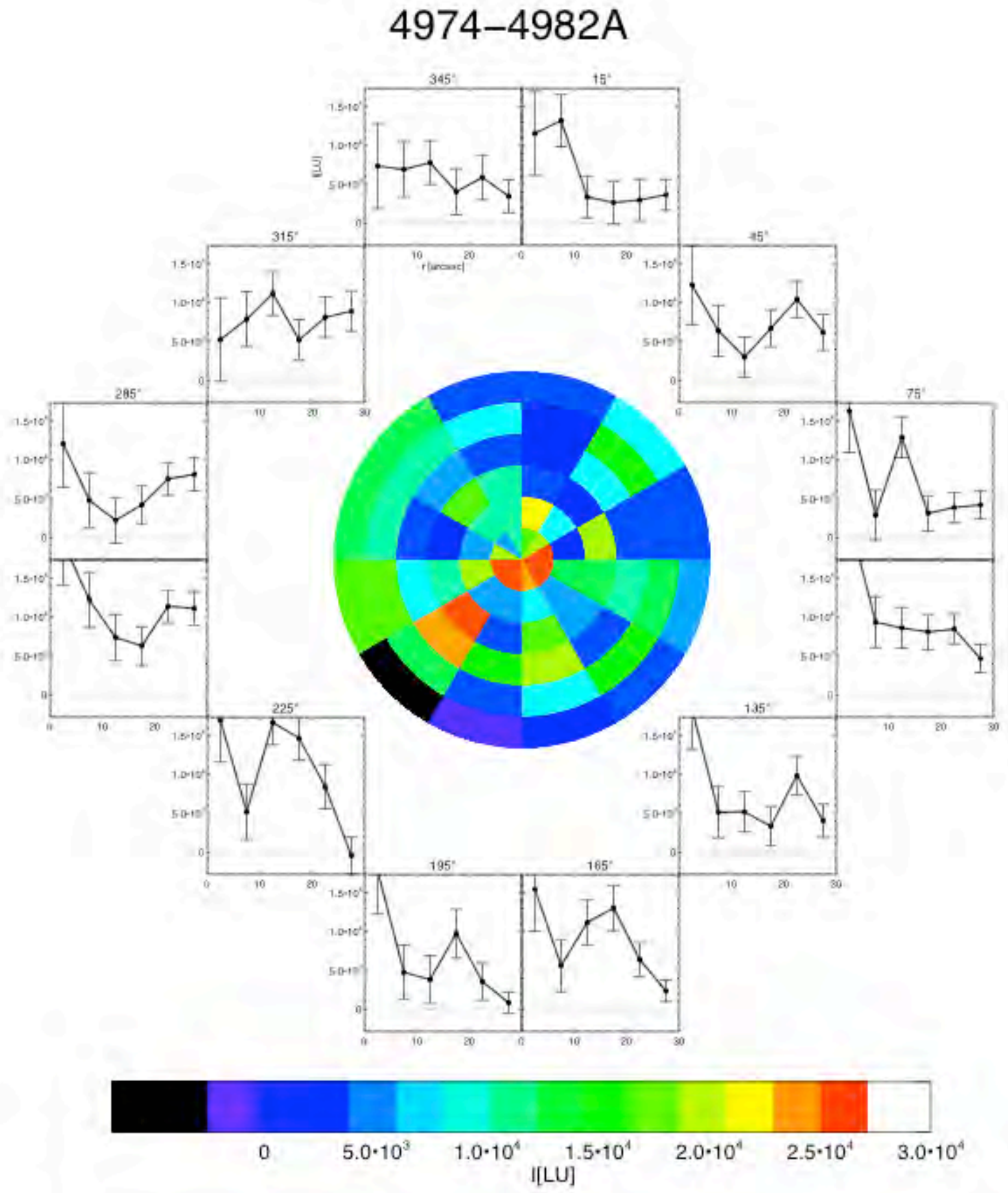

g. Band 4974-4982Å, blue side of LAB2 systemic velocity.

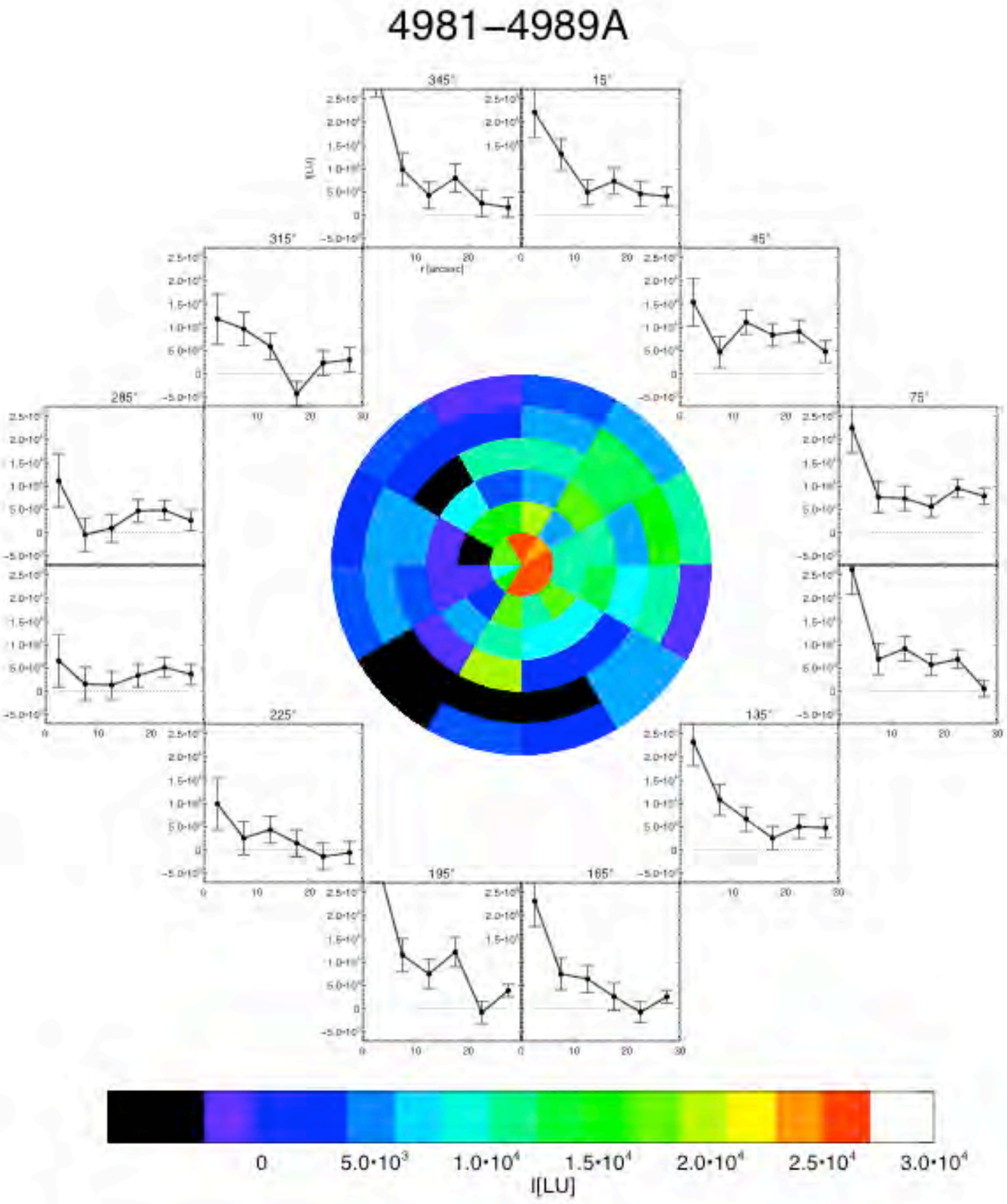

h. Band 4982-4989Å. Red side of LAB2 systemic velocity.

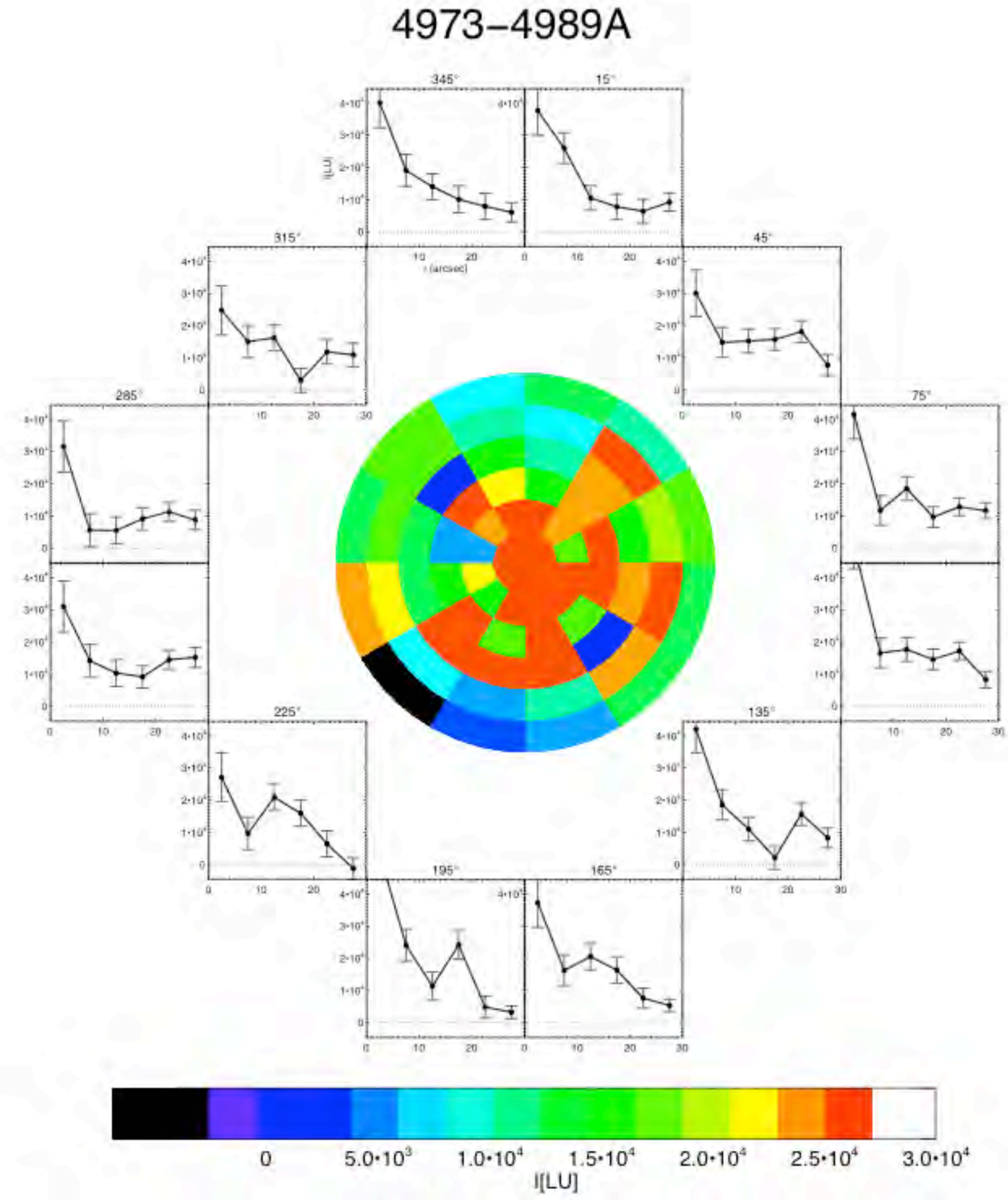

i. Band 4973-4989Å. Blue and red side of LAB2 systemic velocity.

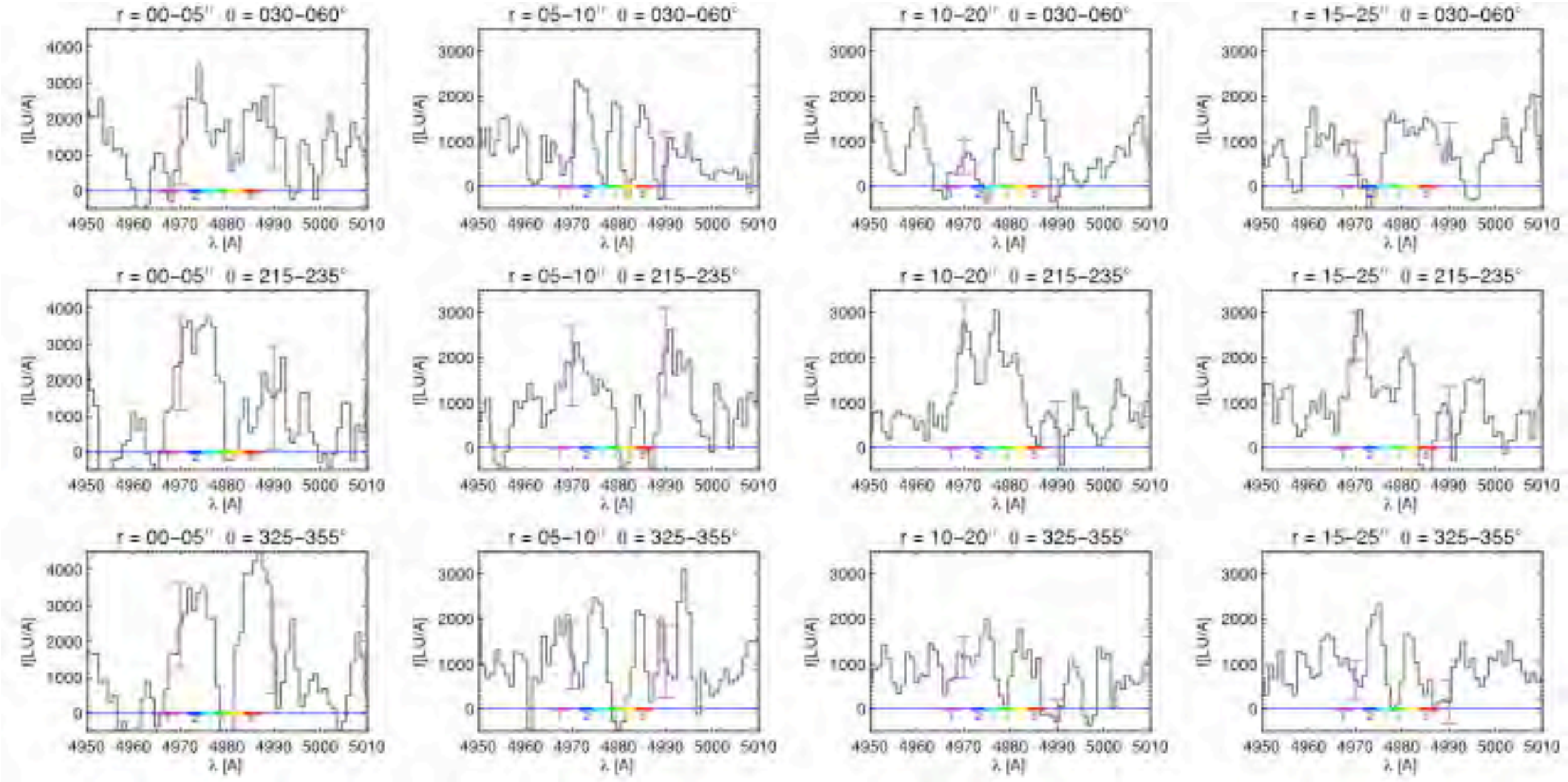

Figure 6. Spectra from radial/azimuthal regions around LAB2. Spectra were extracted from the unsmoothed difference cube in regions defined by the grid in Figure 2, and the error bars shown calculated from the difference variance cube. The spectra shown are smoothed with a 2Å kernel. Each row represents a fixed azimuthal bin, with varying radial intervals. The three azimuthal bins are, from top to bottom, 30-60°, 215-235°, and 325-355°. The four radial bins are, from left to right, 0-5″, 5-10″, 10-20″, and 15-25″. In each of the three rows there appears to be an evolution from the emission features appearing in the LAB2 spectrum to the spectrum in the extended zones. Many single, double, and in some case triple line features are significantly detected. A set of 6 kinematic slices, shown by the numbered colored bars, were selected to produce image slices in Figure 5 and Figure 7.

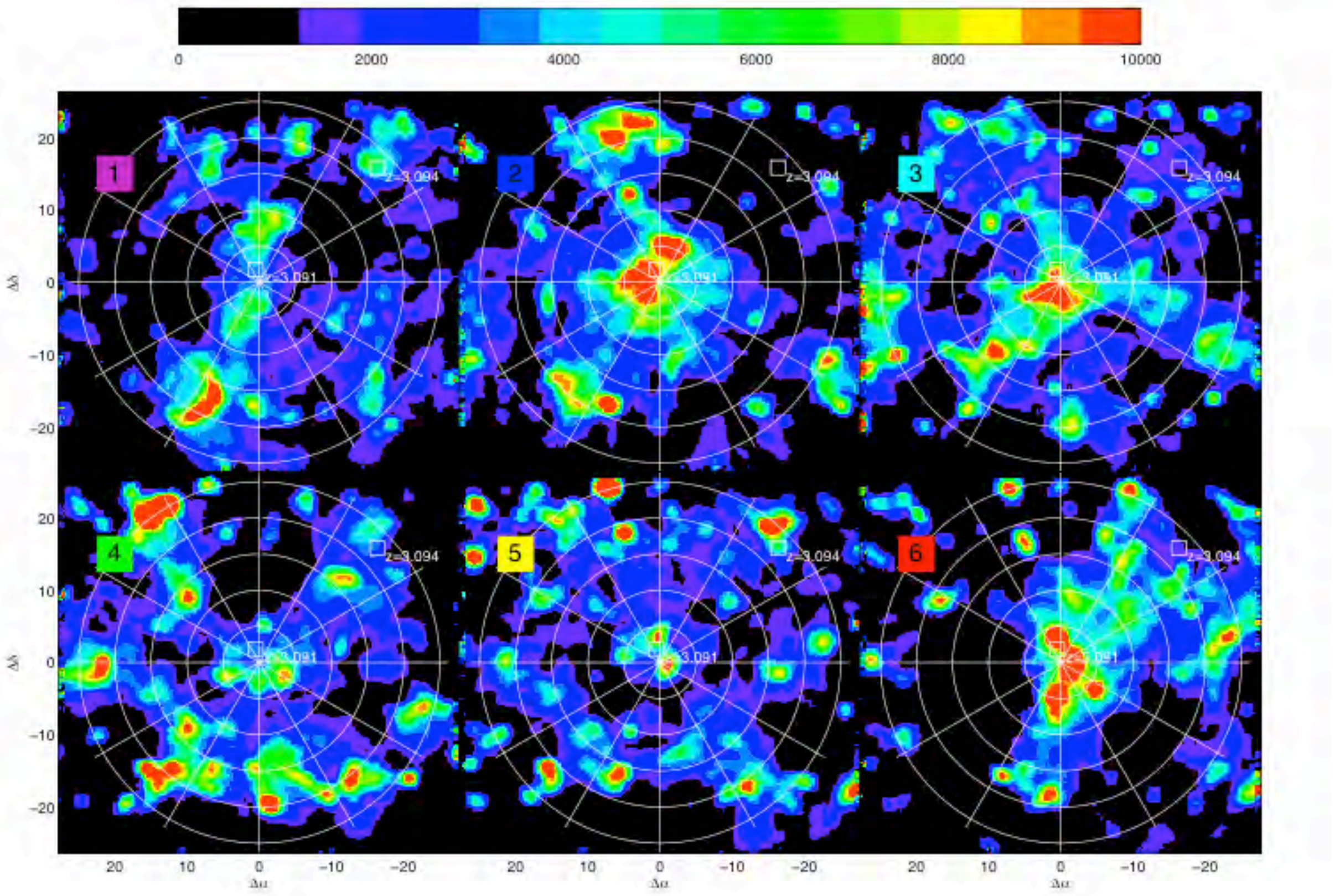

Figure 7. Image slices from adaptively smoothed difference cube. Each panel consists of a 4Å wide slice, centered on 1:4967Å, 2:4973Å, 3:4976Å, 4:4979Å, 5:4982Å, and 6:4985Å.The scale in arcsec is centered on LAB2. Panel 1 shows an extension in the south-east, with similar extensions in panels 2, 3, and 6. One or more extensions can be seen to the west/north-west in panels , 4, and 6. In panels 3 and 6 extensions appear toward the west/south-west. Extensions to the south/south-east, west/north-west, and west/south-west can be seen in the narrow-band image in Figure 2. The color bar gives a linear intensity scale with 10000 LU full range.

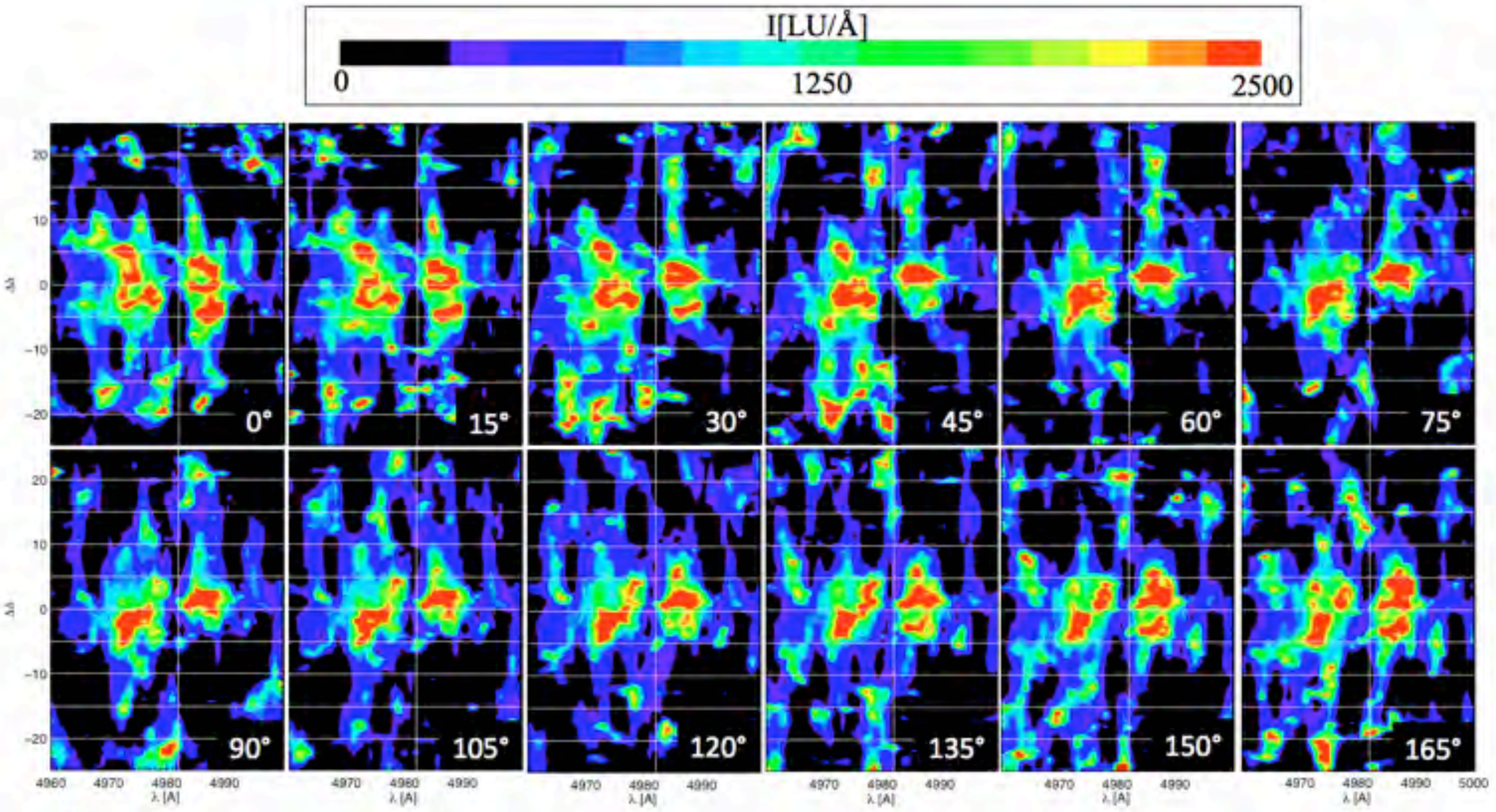

Figure 8. Spectral image plots along radial pseudo-slits. Spectral spatial cuts through adaptively smoothed image cube. Each cut is taken along a radial pseudo-slit 5 arcsec wide at a particular azimuthal angle, separated by 15°. Vertical line shows systemic velocity of LAB2.

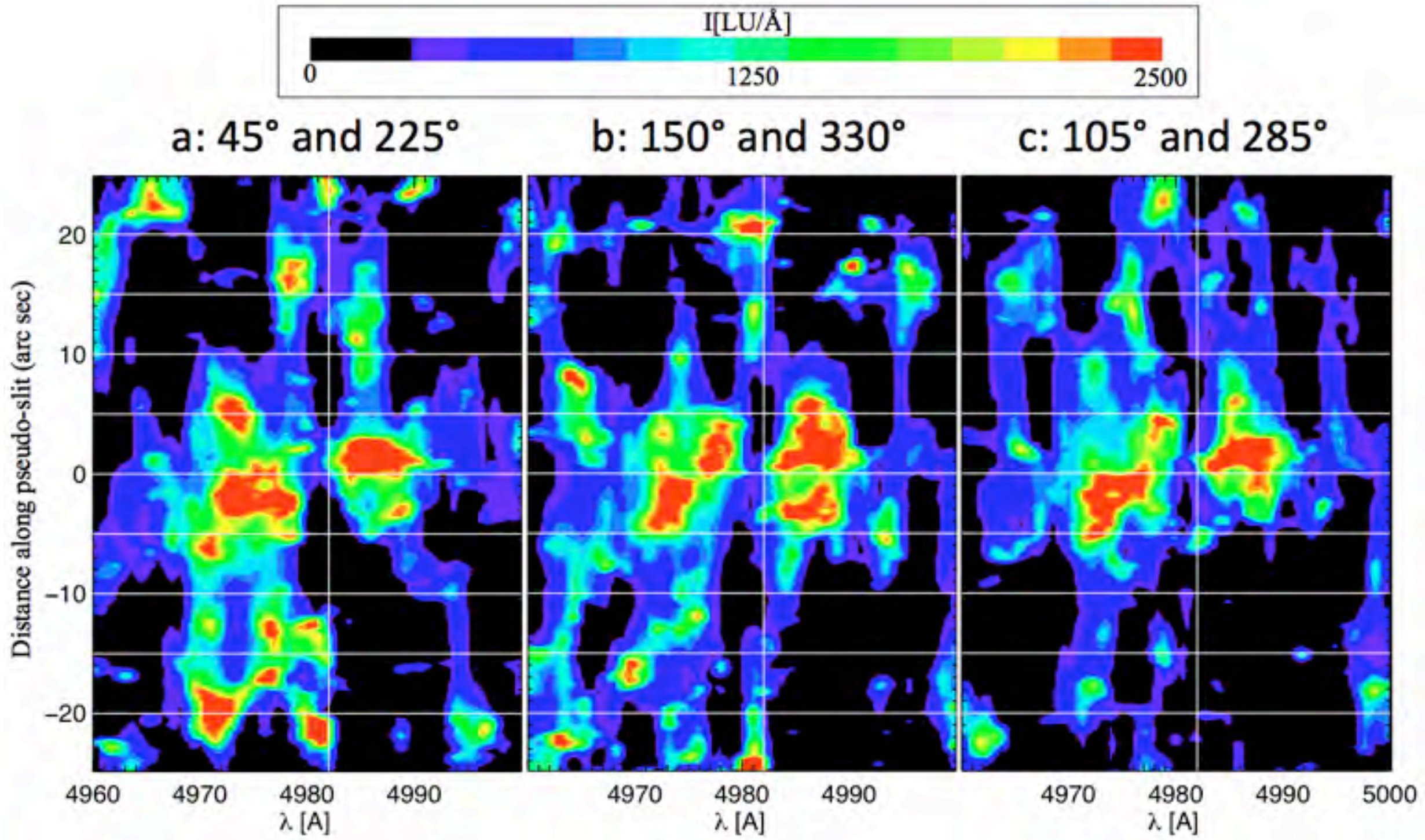

Figure 9. Spectral image plots along radial pseudo-slits. Spectral spatial cuts through adaptively smoothed image cube. Each cut is taken along a radial pseudo-slit 5 arcsec wide at a particular azimuthal angle. Panel a: azimuthal angle 45°. Note that for $\Delta\delta<0$ this corresponds to azimuthal angle 225°. Extended double-peaked emission is detected for $\Delta\delta>0$ with a central wavelength close to that for LAB2 (4982Å shown with white vertical line), with the separation increasing as the blob is approached. More complex, possibly double peaked emission is also present for the azimuth 225°, with a mean wavelength ~4974Å. Panel b: Azimuth 150° ($\Delta\delta>0$) and 330° ($\Delta\delta<0$). Complex double (or triple) peaked emission is seen on the blue side of systemic for azimuth 330°. Weaker emission at the systemic velocity is seen for azimuth 150°. Panel c: Azimuth 105° ($\Delta\delta>0$) and 285° ($\Delta\delta<0$). Azimuth 285° shows only weak extended emission. Azimuth 105° shows weak, possibly double peaked emission with central wavelength ~4972Å. Scale is as in Fig. 3. Note also that some of the velocity spread within LAB2 can be attributed to velocity gradients. Panel (a) and (c) shows that there may be a kinematic axis corresponding to rotation, with angle in the range of 45-105°

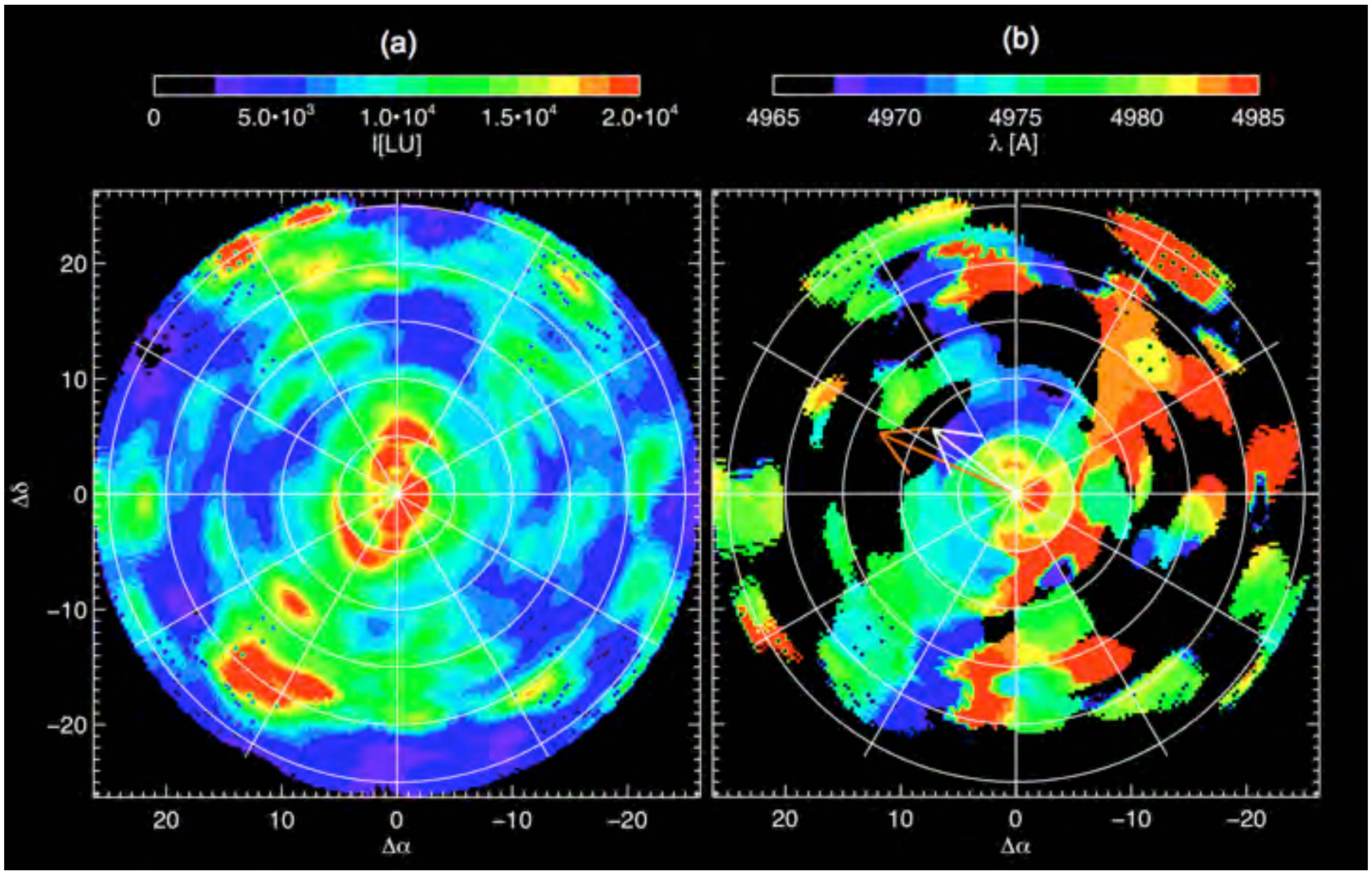

Figure 10. Windowed azimuthal emission plots. These plots are generated to extract extended emission with common but possibly continuously varying kinematics (see text). Panel (a) shows the maximum average intensity in a sliding 12Å wide window. It shows that the extended emission is organized into azimuthal regions 30°-75° (defined as filament 2), 165°-240° (filament 3), and 315°-345° (filament 1). Panel (b) shows the mean wavelength calculated with a weighted sum in the sliding window. Values are shown only when the intensity in panel (a) exceeds 8000LU. Filament 2 is at or slightly red of systemic, filament 1 is blueward of systemic, and filament 3 shows mostly blueshifted emission with some redder components towards the western side. Arrow shows angular momentum vector for the total system (orange) and LAB2 (white, not to scale). Angular momentum is calculated from the unsmoothed difference cube using a mask symmetrical around the spatial and velocity zero, assuming 1 LU/pixel = 1400 $M_\odot$ (obtained by normalizing to the flux from the filaments and the derived gas mass).

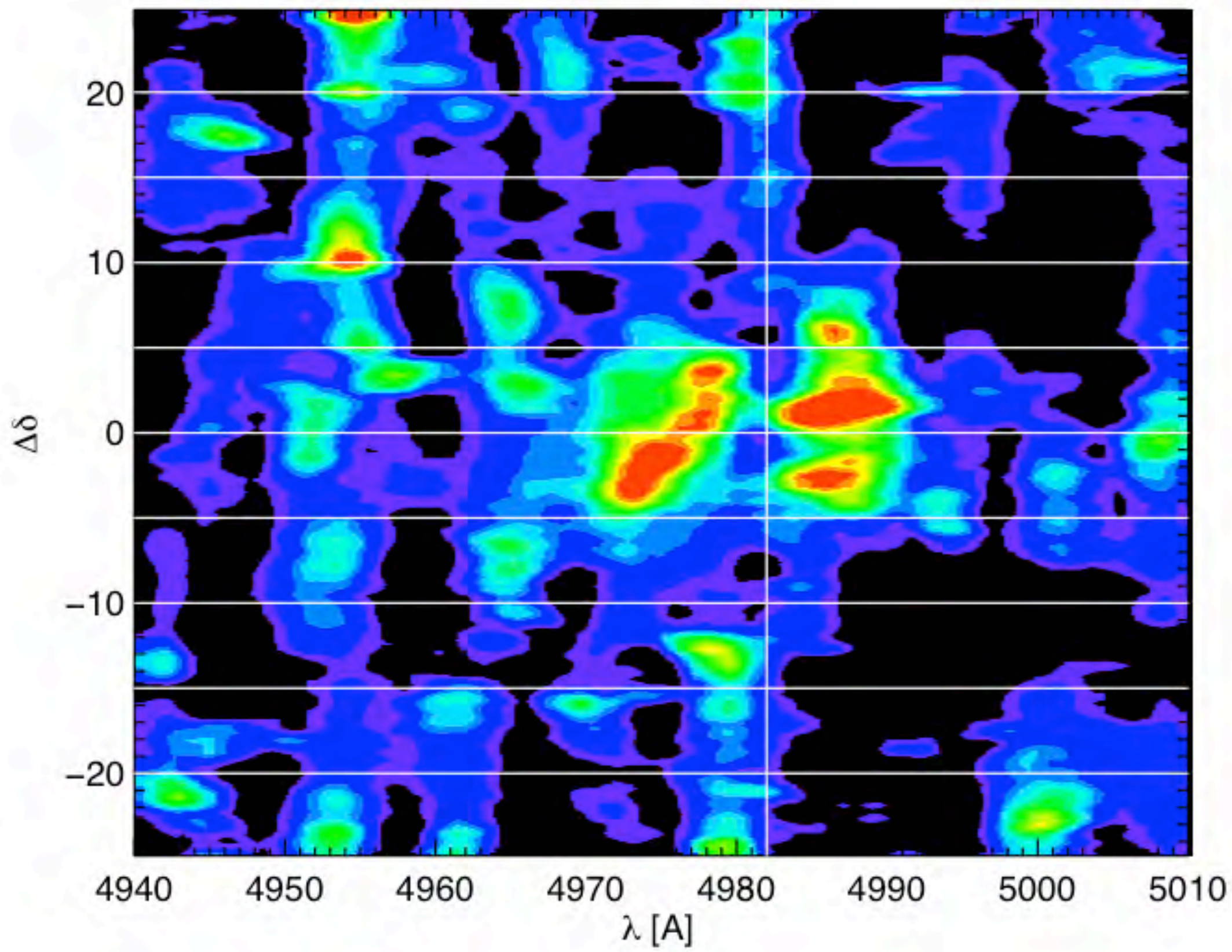

Figure 11. Blue-shifted emission blob near LAB2. This is the spectral image plot for azimuth 135° (Δδ>0) and 315° (Δδ<0). These small blobs visible in the narrow-band image appears at 4953Å, at -1700 km/s with respect to systemic. While it may be unrelated to LAB2 its proximity and appearance on what may be the minor axis of the system is suggestive.

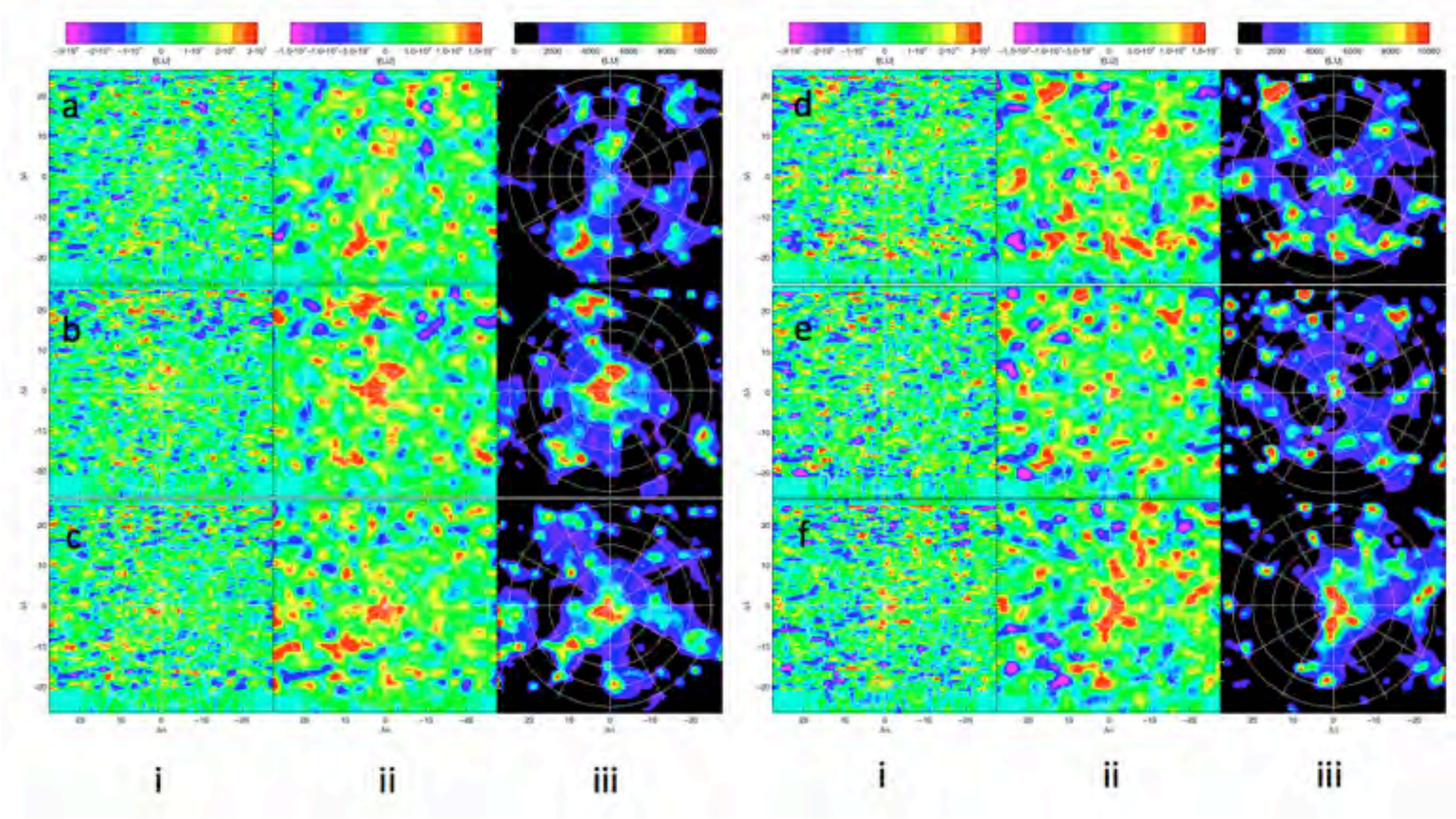

Figure 12. Illustration of raw and smoothed data cube slices, same wavelengths as Figure 7. Letters give wavelengths a. 4967.5Å, b. 4973.5Å, c. 4976.5Å, d. 4979.5Å, e. 4982.5Å, f. 4985.5Å. Small roman numerals give three analysis versions. i. Summed 4Å slice of the raw difference cube with a linear scale shown. In this image 1-sigma noise is ~12,000 LU. ii. In the middle panel we show a version of the left image smoothed with a 9-pixel (2.5 arcsec) boxcar, which has a 1-sigma noise of ~4000 LU. iii. The right panel is identical to Figure 7. And is adaptively smoothed, using low-to-high wavelength smoothing sequence (see text), and a signal-to-noise threshold of S/N>2.5. We note that the correspondence between the central and right-hand panels is not one-to-one. This is because the smoothing algorithm uses a more complex procedure (see text) that includes a pass with 2Å and 8Å smoothing.

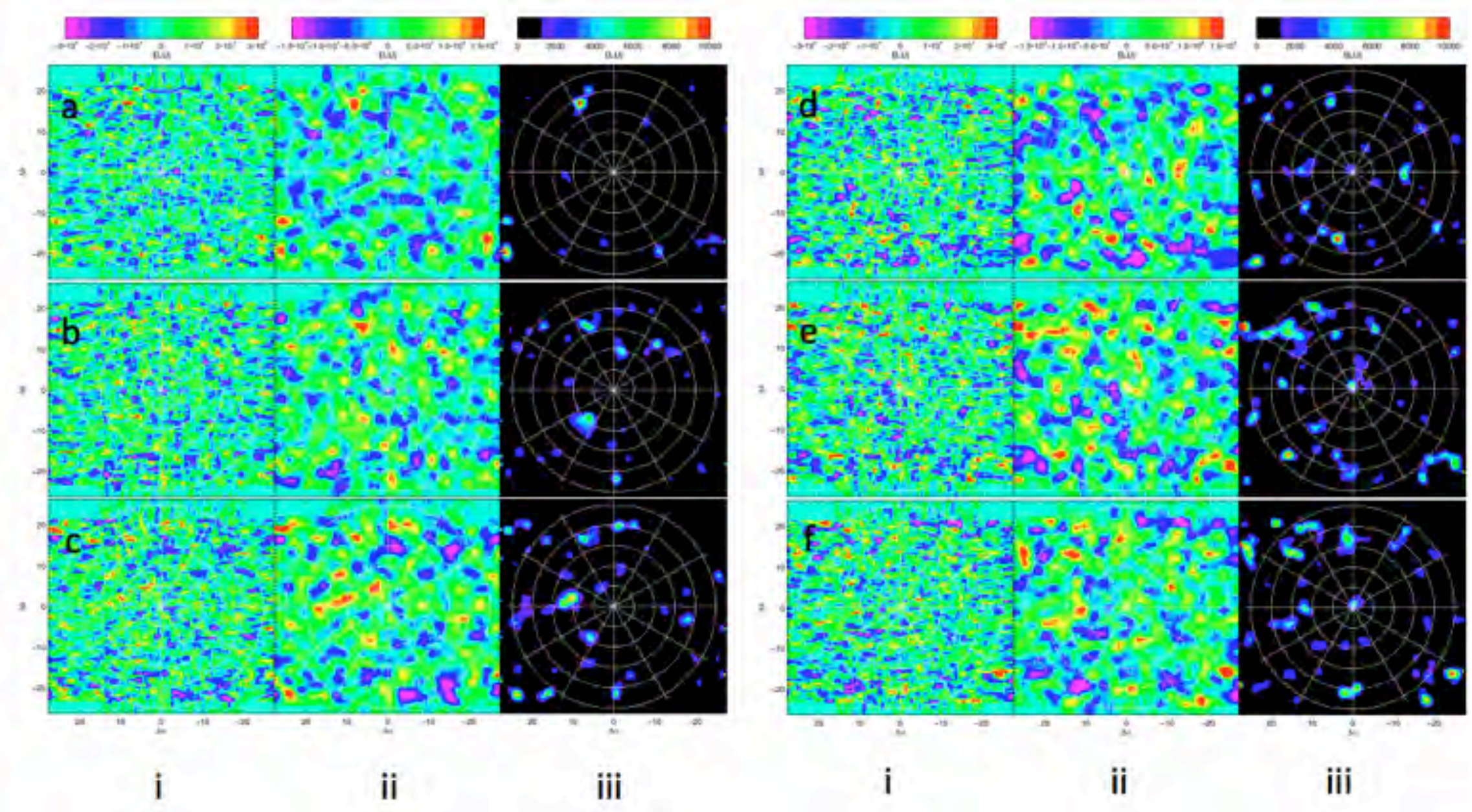

Figure 13. Panels showing noise simulation, in the same format as Figure 12. We simulated a noise field with no emission to compare to Figure 12. Letters give slice wavelengths identical to Figure 12. a. 4967.5Å, b. 4973.5Å, c. 4976.5Å, d. 4979.5Å, e. 4982.5Å, f. 4985.5Å. Small roman numerals give three analysis versions. i. Summed 4Å slice of the raw difference cube with a linear scale shown. In this image 1-sigma noise is ~12,000 LU. ii. In the middle panel we show a version of the left image smoothed with a 9-pixel (2.5 arcsec) boxcar, which has a 1-sigma noise of ~4000 LU. iii. The right panel is identical to Figure 7. And is adaptively smoothed, using low-to-high wavelength smoothing sequence (see text), and a signal-to-noise threshold of S/N>2.5. The noise level varies somewhat from slice to slice because of the presence of Na emission lines.

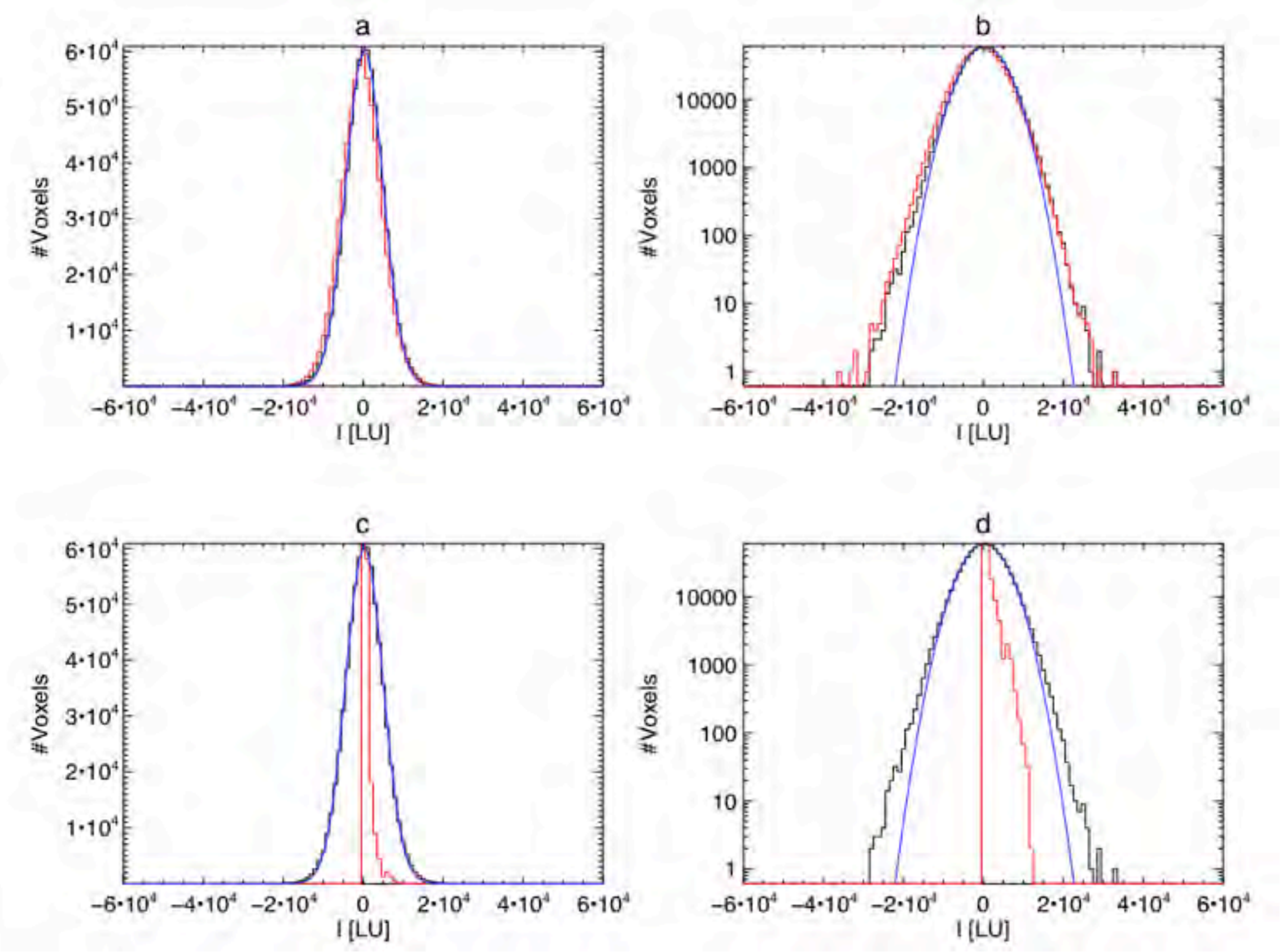

Figure 14. Histogram of voxel values in difference cubes. a. Data (black), noise simulation (red), fit to data with single Gaussian (blue). b. Same as a. on log scale. C. Data (black), 3DASL (red), fit to data with single Gaussian (blue). d. Same as c. on log scale.

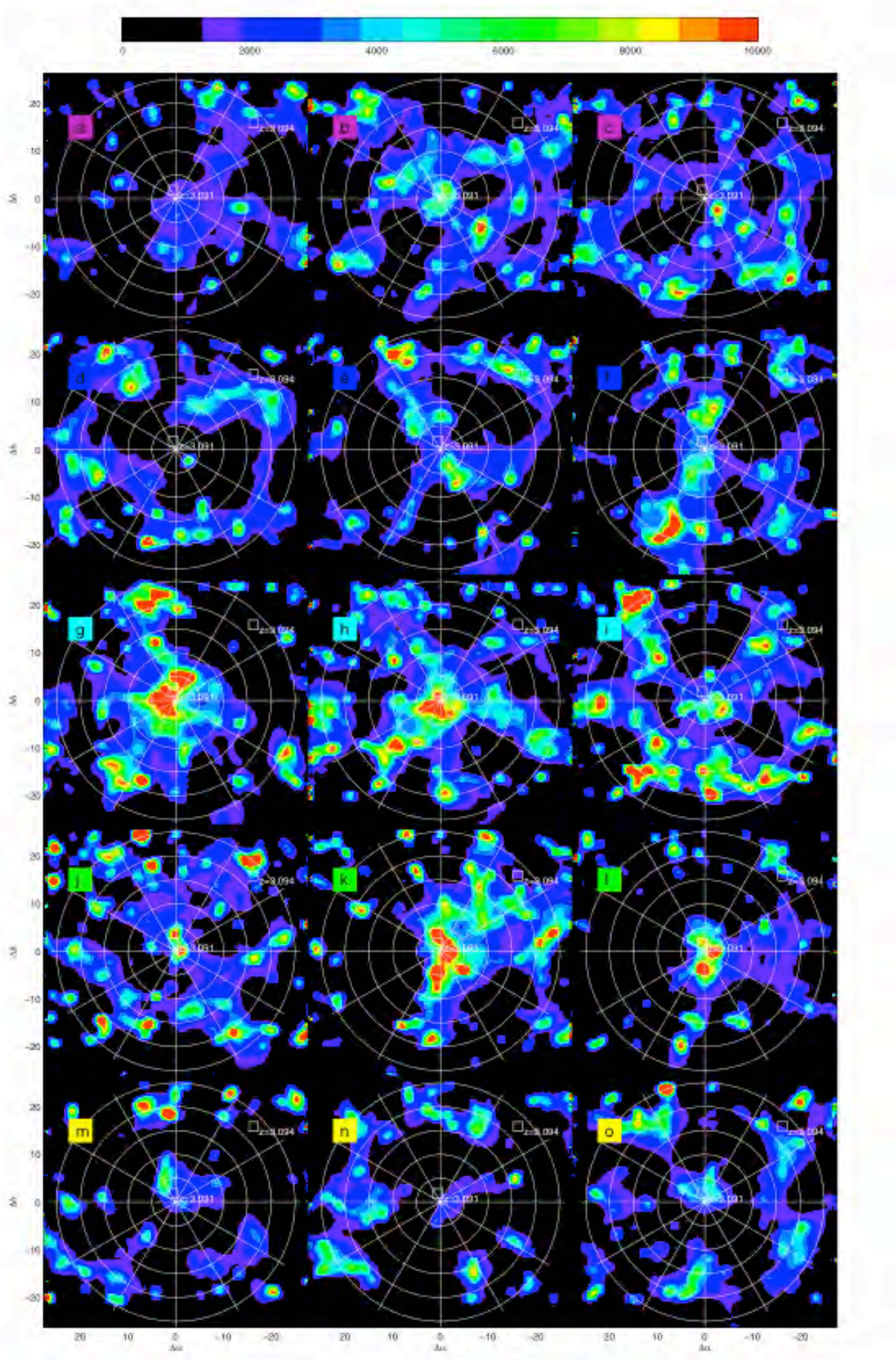

Figure 15. Extended version of image slices shown in Figure 7. a. 4948Å, b. 4952Å, c. 4956Å, d. 4960Å, e. 4964Å, f. 4967.5Å, g. 4973.5Å, h. 4976.5Å, i. 4979.5Å, j. 4982.5Å, k. 4985.5Å, l. 4990Å, m. 4994Å, n. 4998Å, o. 5002Å.

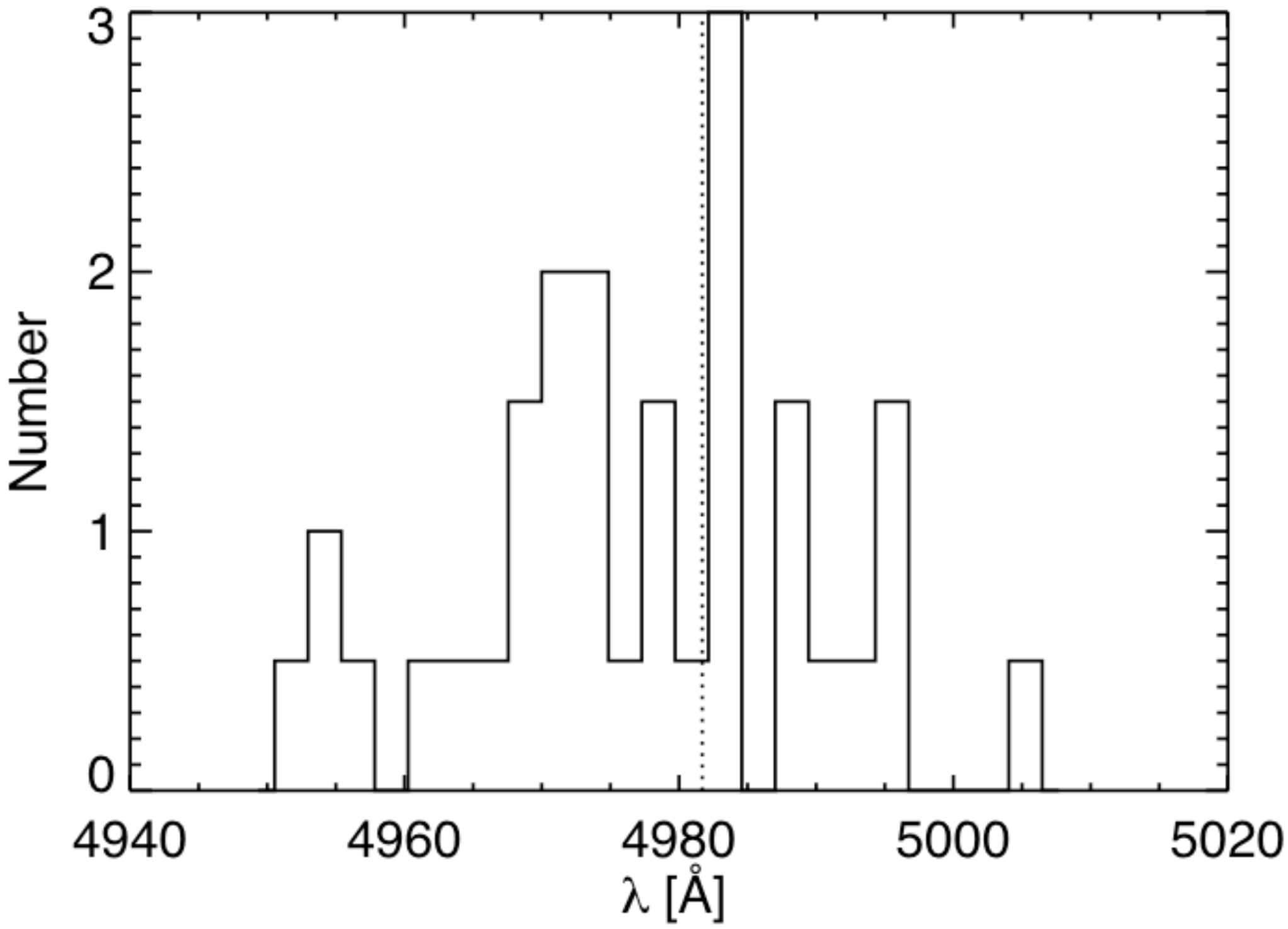

Figure 16. Distribution of galaxies in SSA22 area. Significant structure is present from 4950Å to 5006Å.

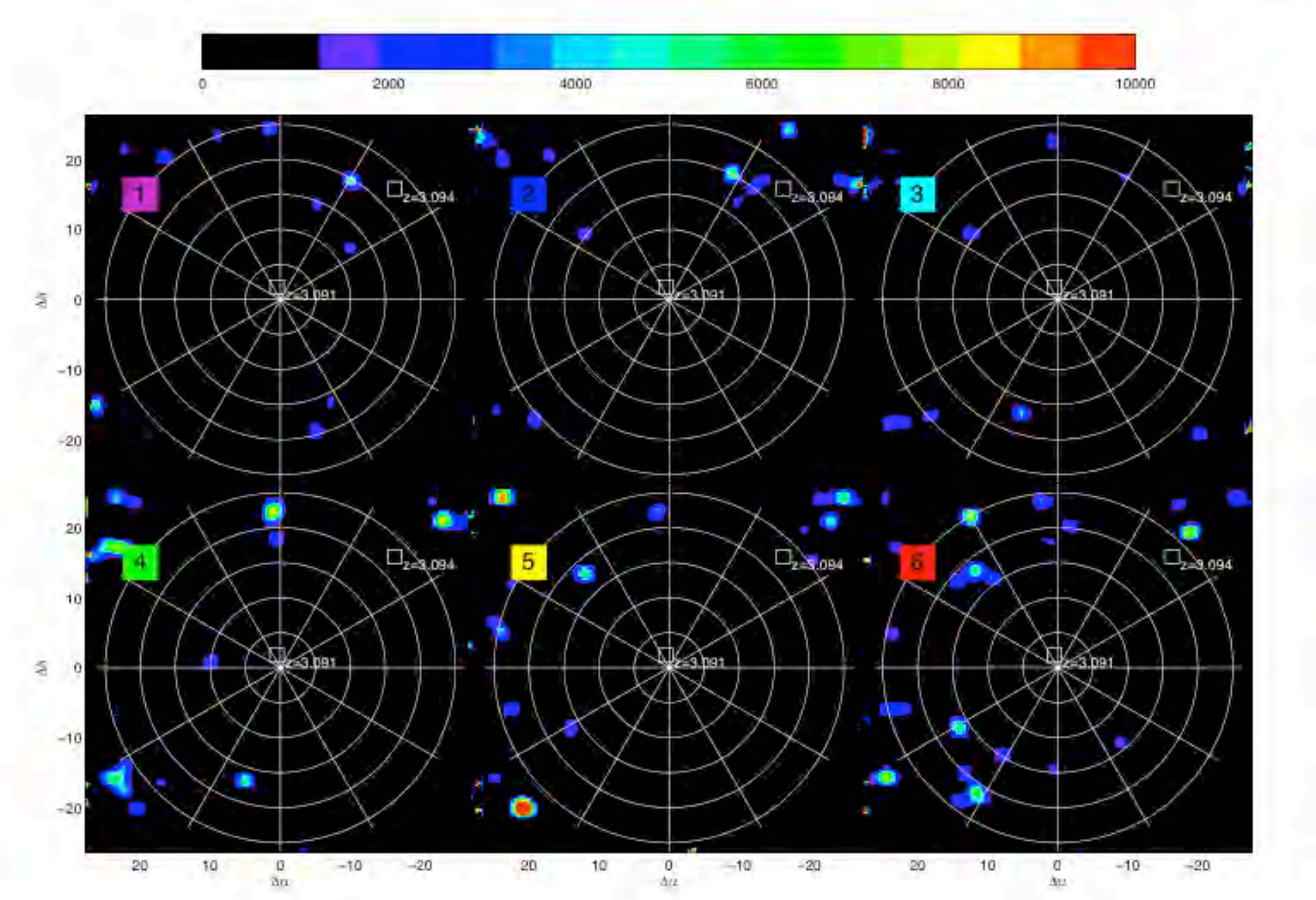

Figure 17. 3DASL run on inverted data cube (data cube multiplied by -1). Six slice shown in format identical to Figure 7.

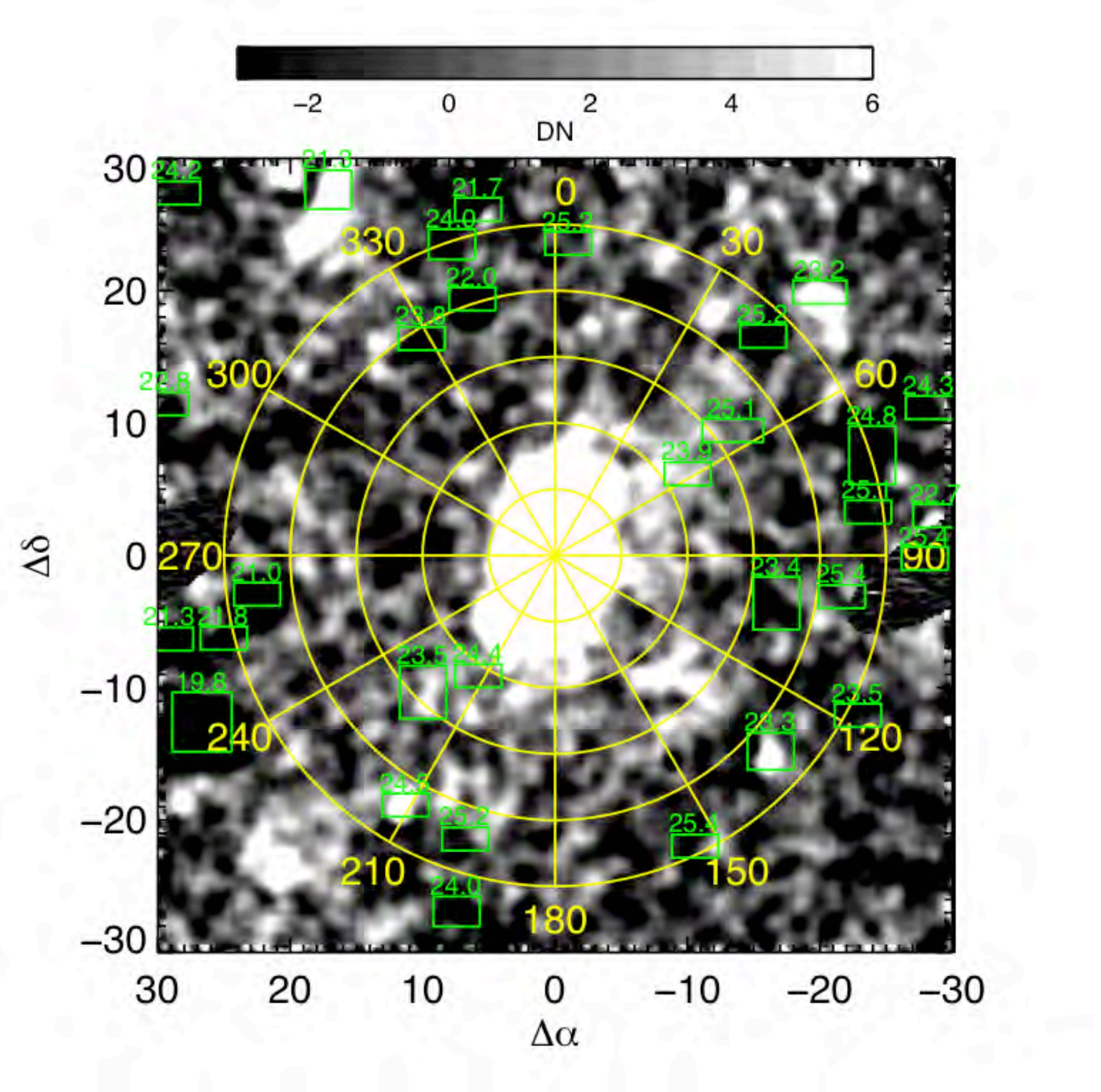

Figure 18. Location of compact sources subtracted from data cube. Compact sources were detected in broad-band images. Based on the size and magnitude of each source, a subtraction region was generated, as shown superimposed on the narrow-band image in the green boxes labeled by the approximate source magnitude (AB, V-band). A median sky value was determined in adjacent spaxels at each wavelength and used to replace the spaxels in the source box location.

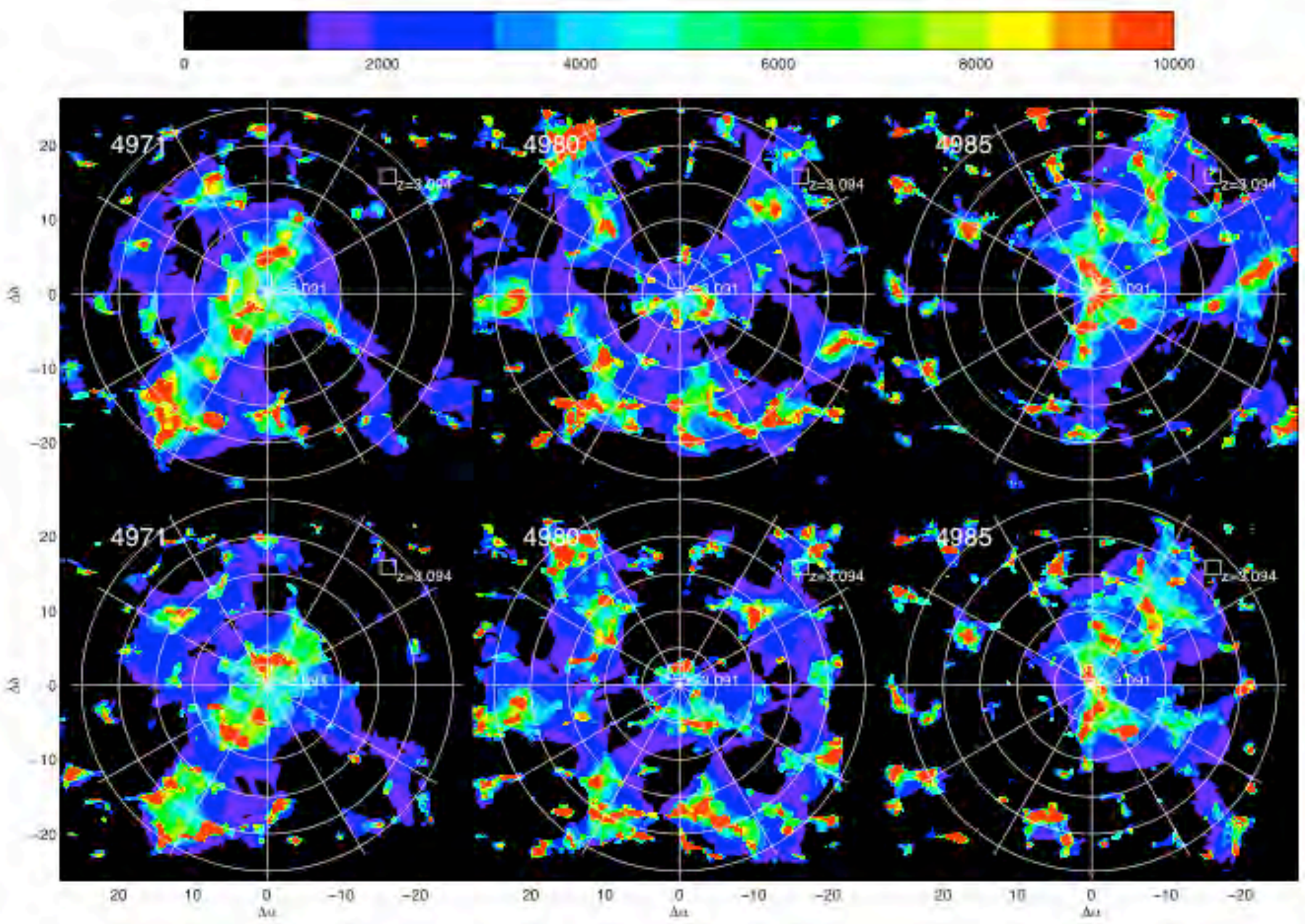

Figure 19. Effect of compact source subtraction on images. This figure shows the smoothed images summed in 4Å slices for (left to right) 4971Å, 4980Å, and 4985Å. The top row shows shows after subtraction of sources brighter than V=25.5. All images use default algorithm with threshold S/N>2.5, while the bottom row shows the result with no source extraction.

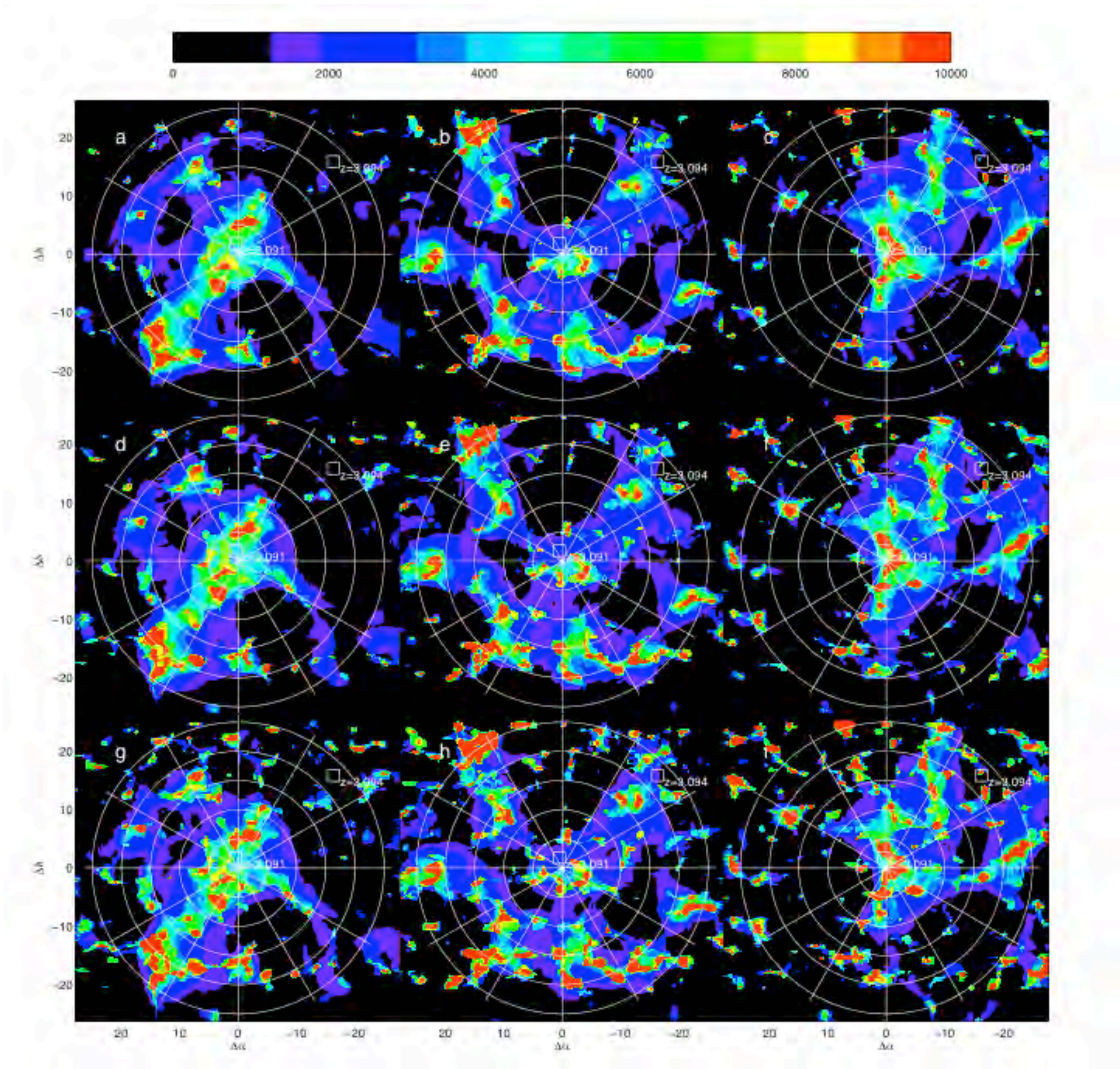

Figure 20. Effect of signal-to-noise threshold. This figure shows the smoothed images summed in 4Å slices for (left to right) 4971Å, 4980Å, and 4985Å. From top to bottom we show signal-to-noise threshold of S/N>3.0, S/N>2.5, and S/N>2.0. As the threshold is lowered, noise spikes become more frequent, but are generally compact. Emission regions with a gradient in emission level change and shift somewhat as more compact regions rise above the threshold and are removed from the more heavily smoothed contributions.

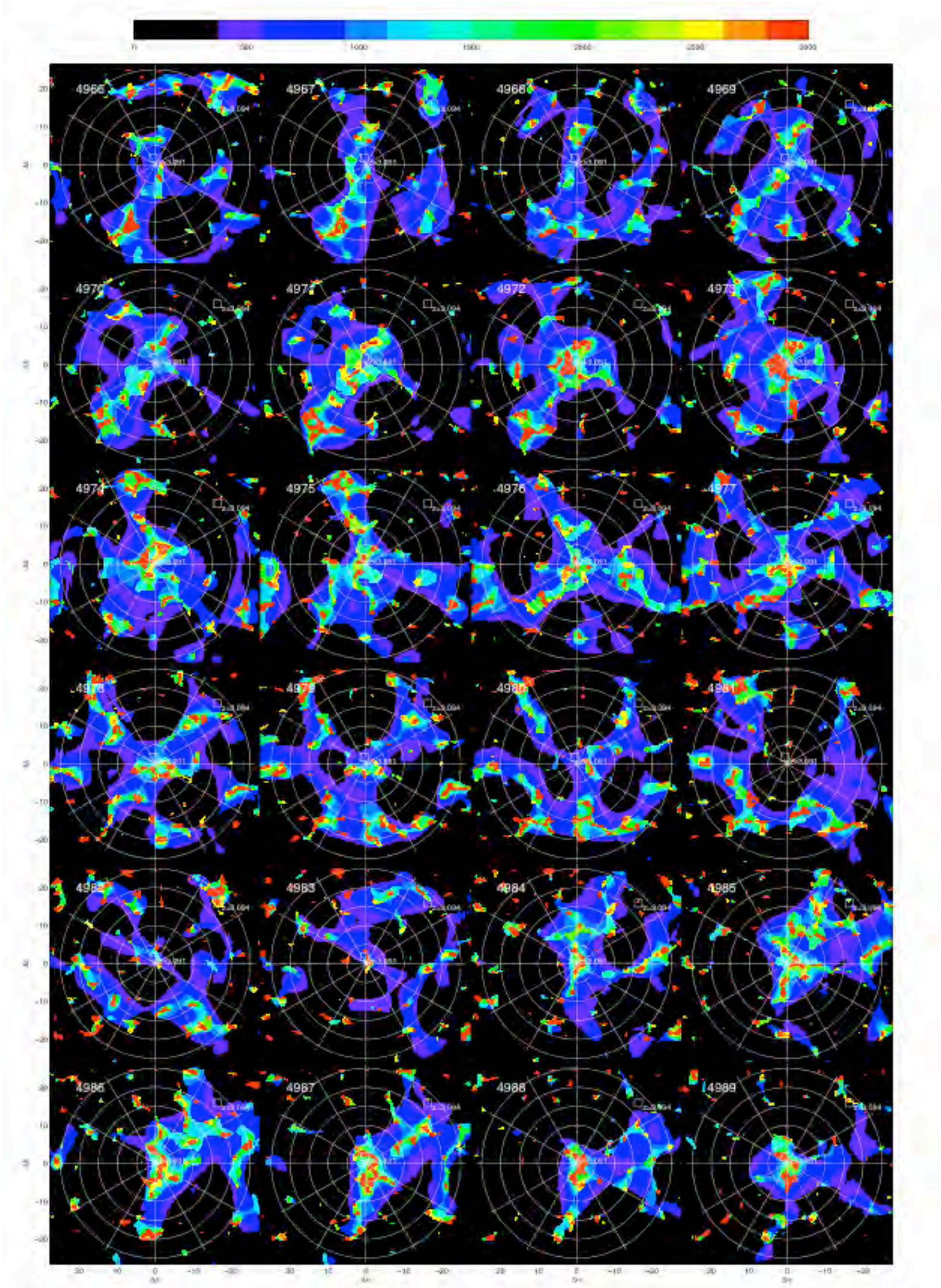

Figure 21. Detection in multiple slices. Smoothed 1Å slices using 3DASL covering band 4968-4987Å. S/N threshold is 2.5. Many features are detected in multiple slices. Filament 1 (see text for definitions) is detected in at least 4 slices (4972-4980Å). Filament 2 is detected in 8 slices [4977-4980Å, 4984-4987Å]. Filament 3 is detected in 7 slices (4966-4972Å).

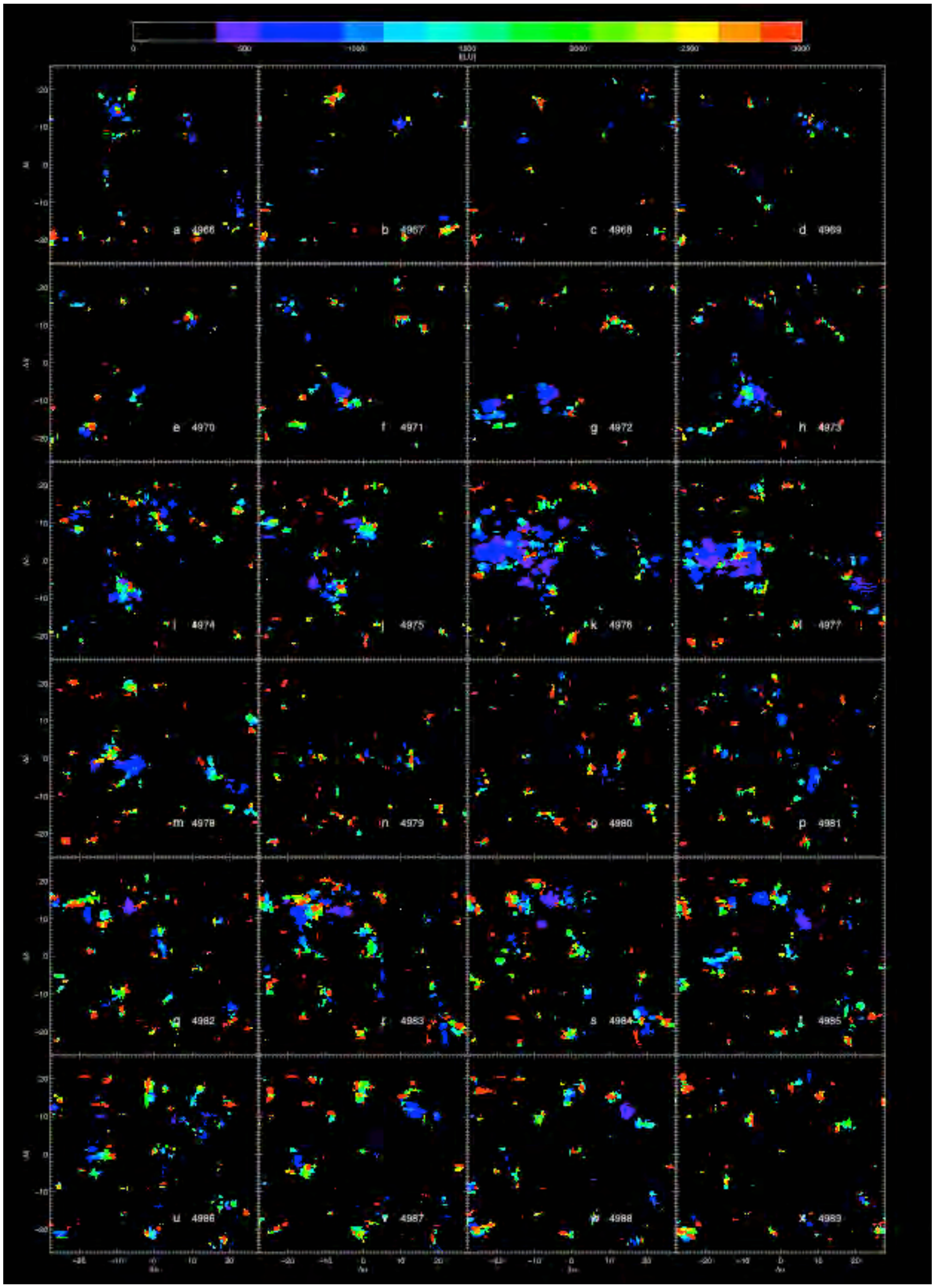

Figure 22. Detection in multiple slices in pure noise data cube. Smoothed 1Å slices using 3DASL covering band 4968-4987Å, with pure noise based on simulation as shown in Figure 13. S/N threshold is 2.5. In general spurious noise features appear in at most 2 wavelength slices.

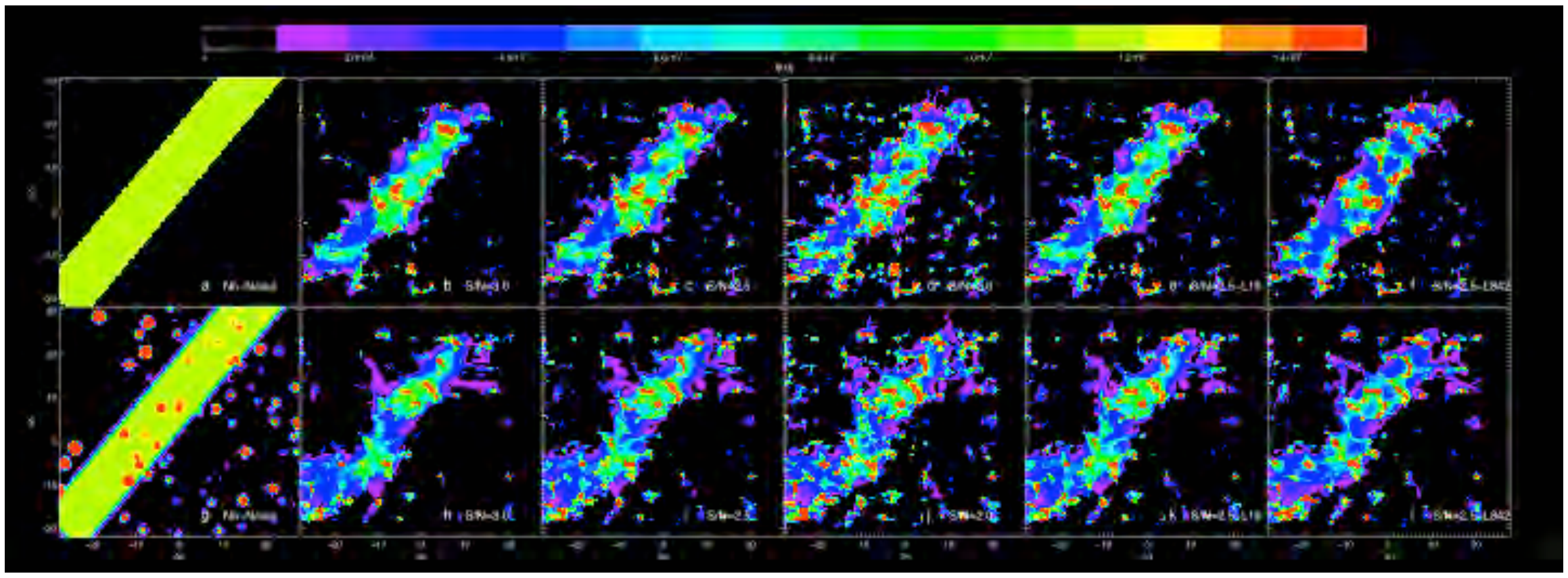

Figure 23. Simple filament simulations. Simulation of a single uniform filament, total flux 15000LU in double-peaked optically thick Lyα line profile with peak separation 3.5Å, velocity dispersion 60 km/s, filament width 12 arcsec, length 60 arcsec, position angle 40°, convolved with a 1.2 arcsec FWHM seeing disk. In the top row, no continuum sources are added. In the bottom row, continuum sources present in the LAB2 field are added, distributed in intensity from V=20.7 to 26.0, and sources are subtracted from the processed data cubes. Images are 8Å sums centered on the line center. From left to right we show: a. the input filament; b. three choices of signal-to-noise threshold: S/N>3.0, S/N>2.5, S/N>2.0 with the default algorithm; e. an additional 16Å smooth step is added, S/N>2.5; f. the 2Å, 4Å, 8Å smoothing sequence is reversed to 8Å, 4Å, 2Å.

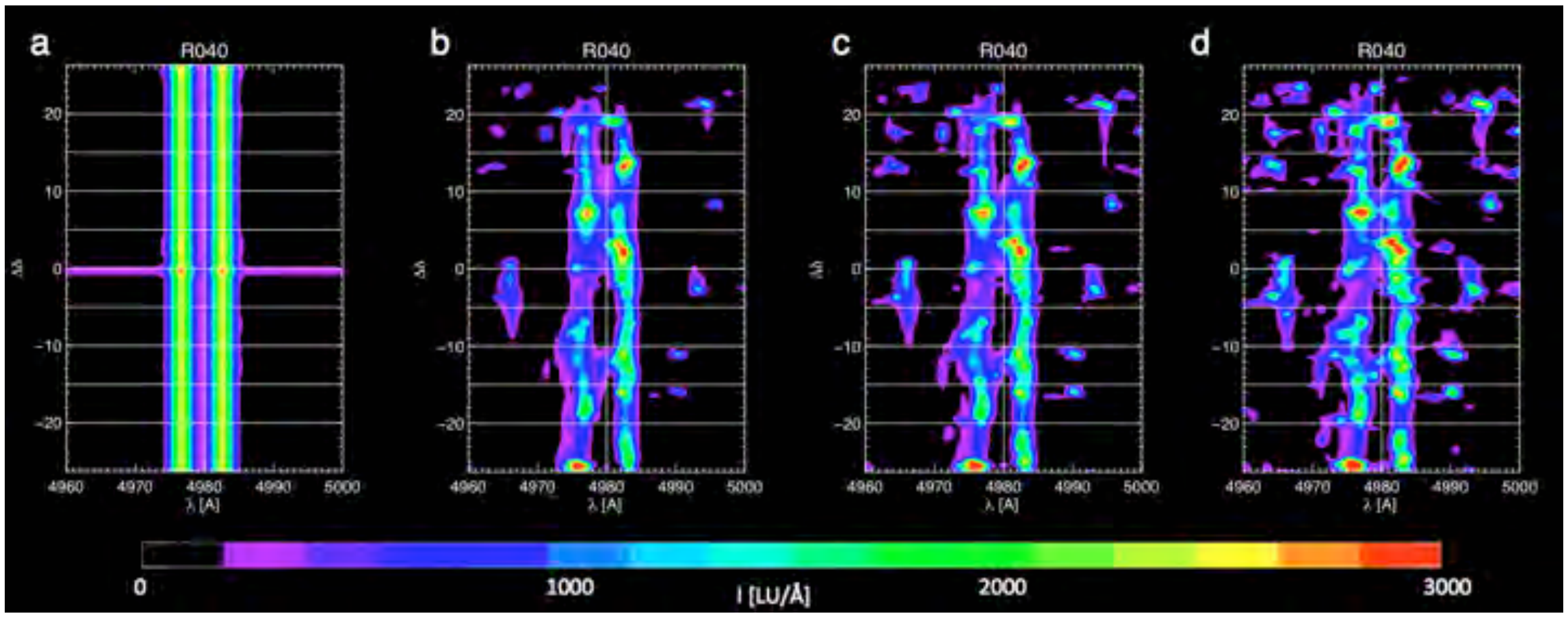

Figure 24. Simple filament simulations, spectral-image plots. We show here the spectral-image plots taken with pseudo-slit oriented at 40°, with 5 arcsec width. a. simulation input cube; b. smoothed data cube, S/N>3.0; c. ; c. smoothed data cube, S/N>2.5; d. smoothed data cube, S/N>2.0. Sources are added in the input image and subtracted from the data cube.

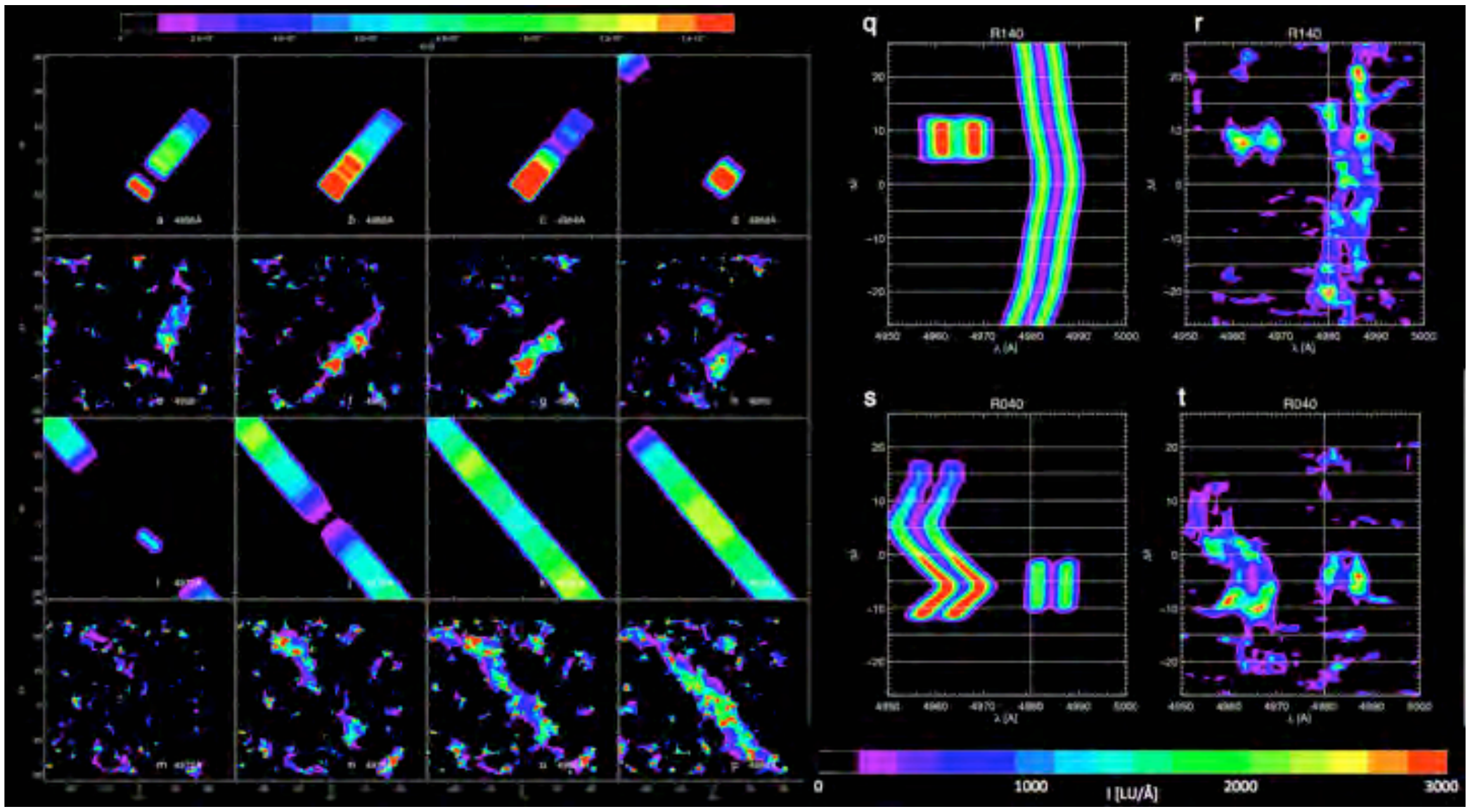

Figure 25. Simple two-filament simulation with velocity gradients. Two filaments have been simulated, both with double peaked profiles as in the previous one filament simulation. A short filament is centered on 4960Å, and a long filament at 4982Å. Panels a-d and i-l give successive slices of the simulation, while e-h and m-p give the corresponding smoothed data cube slices, with S/N>2.5 threshold. Panels q and s give the spectral image slices of the simulation input, oriented along the long and short filaments respectively. Note that velocity gradients yielding emission center variations of 5-10Å are present in the filaments. Panels r and t give the output spectral image slices. The noisy output filaments still reproduce the input velocity profiles, albeit with some modest discrepencies.

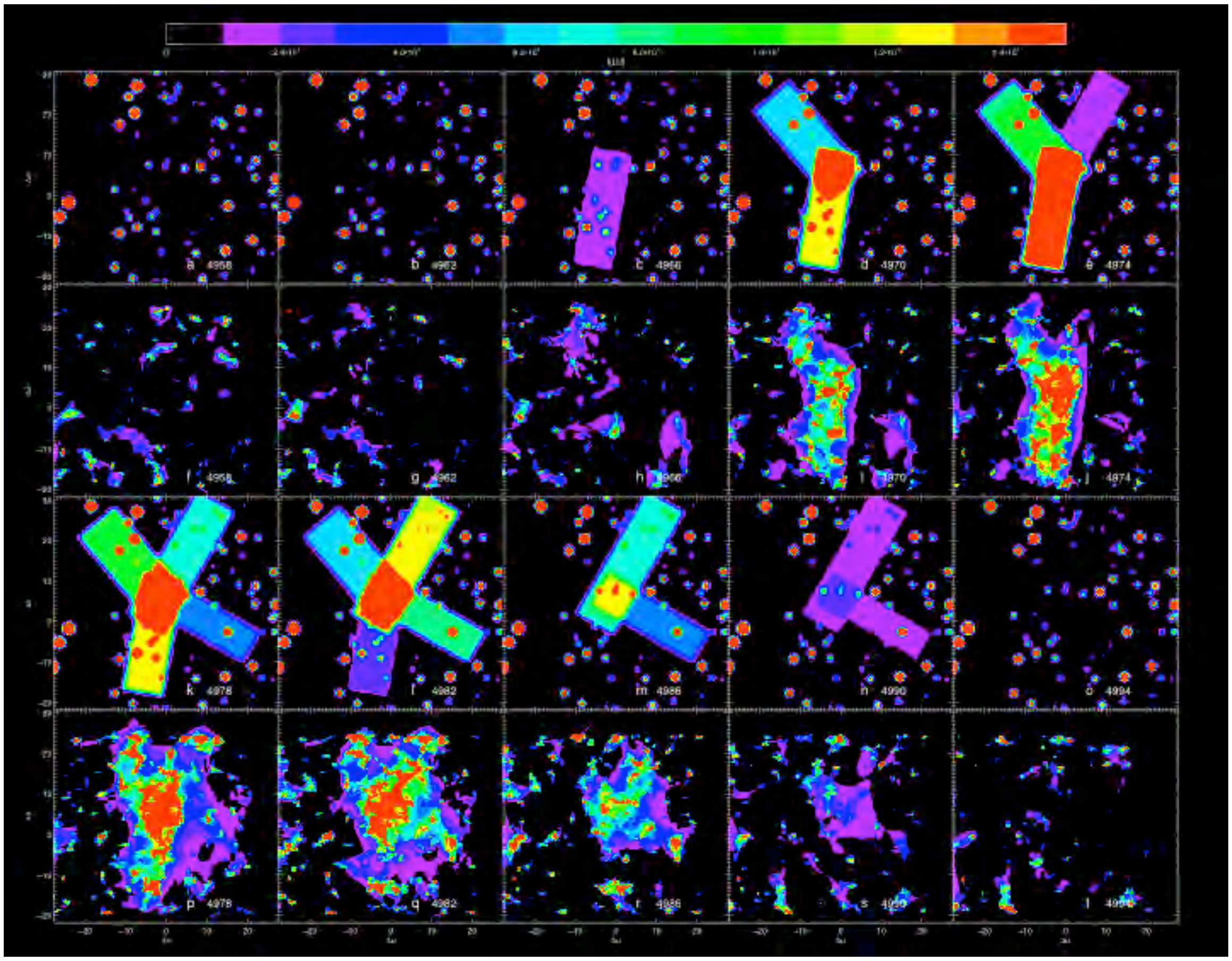

Figure 26. Representative filament simulations—resolving velocity components. This simulation is of 4 filaments and sources roughly similar in position and flux to that observed around LAB2, but with a much simpler kinematic structure. In the simulation the southern filament is centered at 4974Å, while the other three are centered at 4982Å. Proceeding horizontally from left-to-right and from top to bottom each pair of images shows a no-noise version, and the noisy, smoothed 8Å wide slice. Slices are shown every 4Å from 4958Å to 4990Å. All images were source subtracted and smoothed with a threshold S/N>2.5.

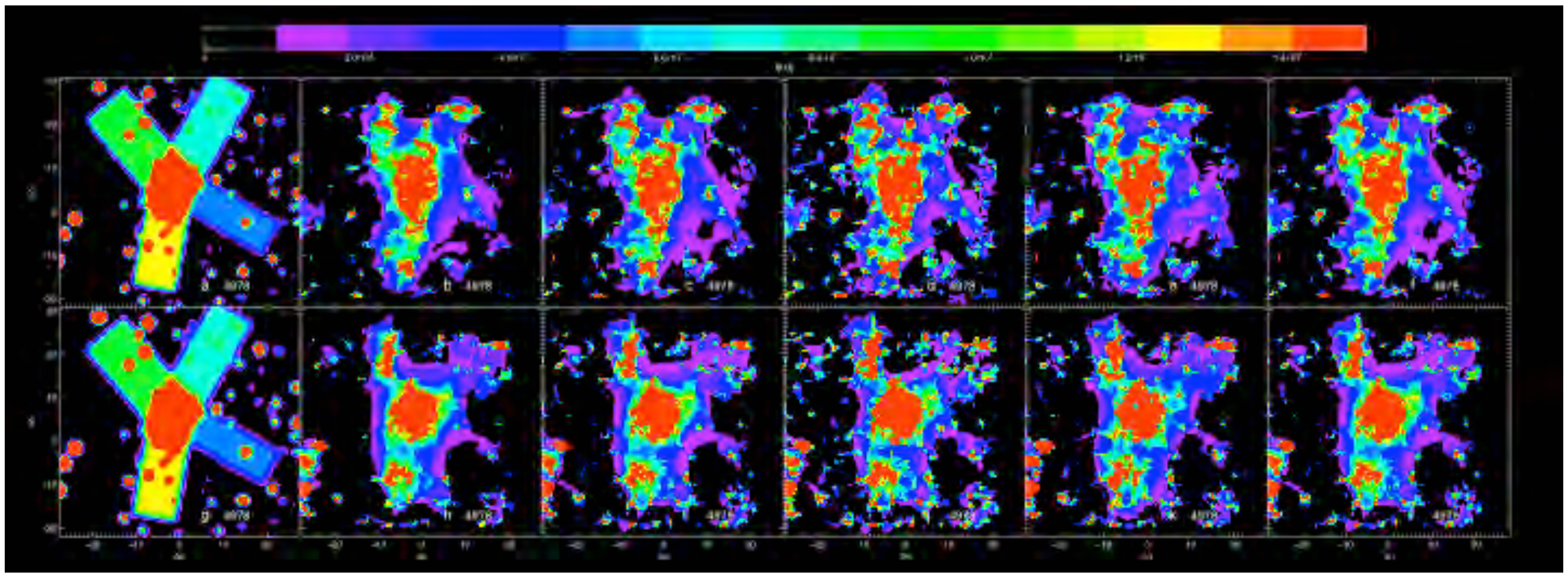

Figure 27. Representative filament simulations—effect of source subtraction and S/N threshold. This simulation is of 4 filaments and sources roughly similar in position and flux to that observed around LAB2, as shown in Figure 26. From left to right, simulation, smoothed S/N>3.0, S/N>2.5, S/N>2.0, S/N>2.5 with 16Å smoothing step, and reversed smoothing (8Å → 4Å → 2Å). Top row has sources subtracted, bottom row has now source subtraction.

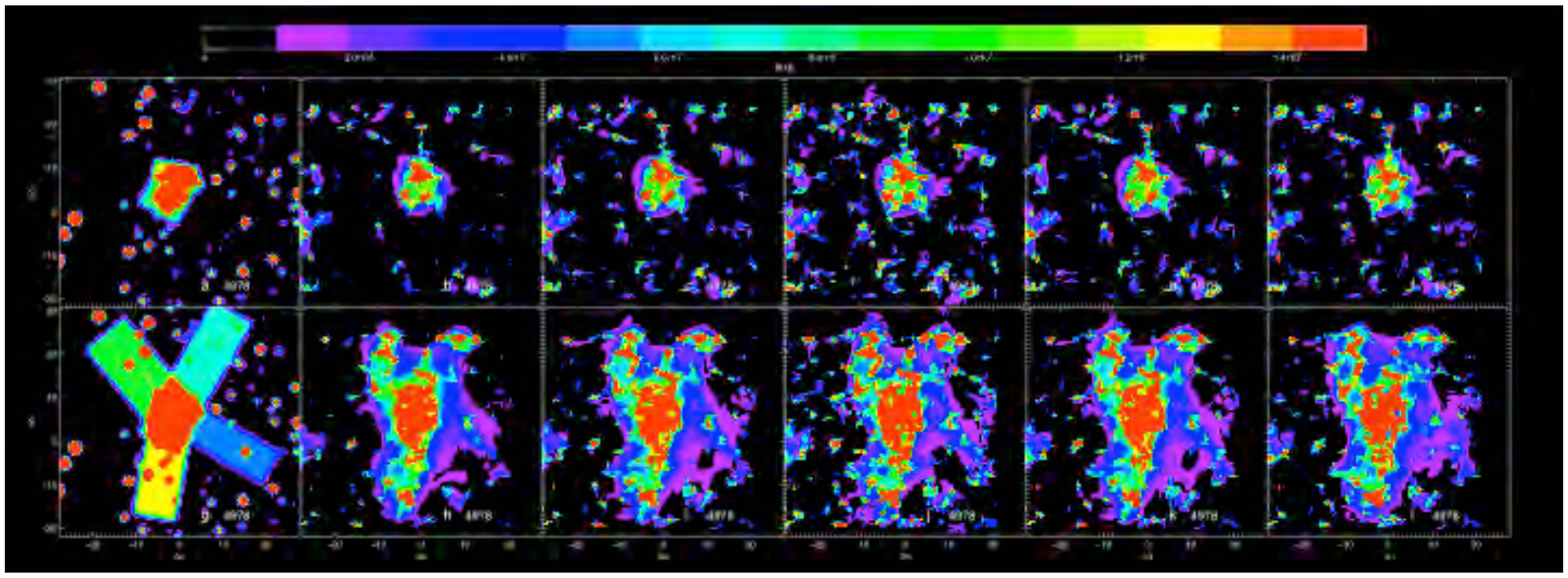

Figure 28. Comparison of simulated blob without extended filaments and blob with extended filaments, 8Å slice, centered on 4978Å. Top row shows blob made up of short filaments, bottom blob with extended filaments. From left to right, simulation, smoothed S/N>3.0, S/N>2.5, S/N>2.0, S/N>2.5 with 16Å smoothing step, and reversed smoothing (8Å → 4Å → 2Å). Both simulations have sources subtracted.

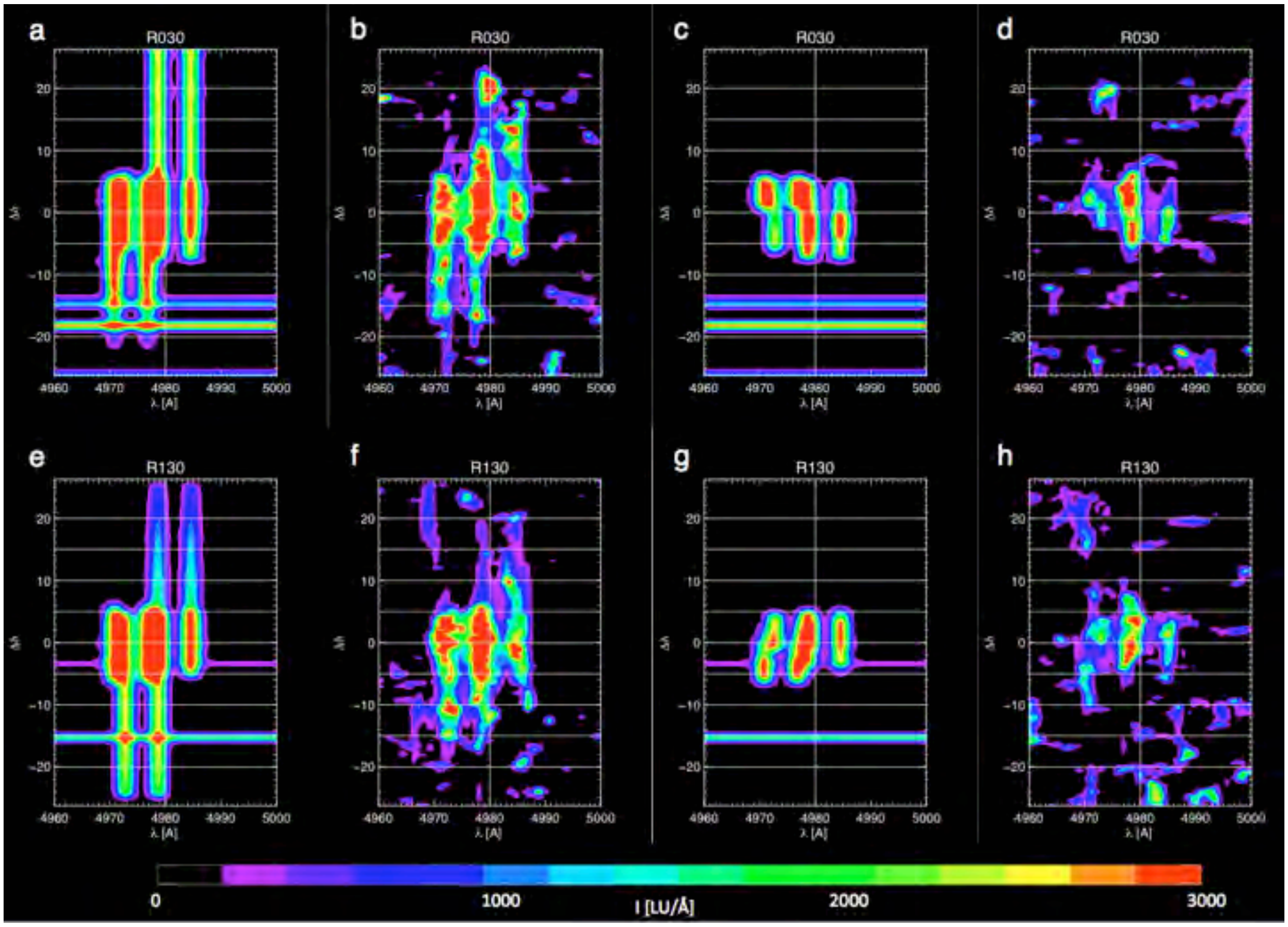

Figure 29. Comparison of simulated blob without extended filaments and blob with extended filaments, spectral-image plots. Panels a-d, azimuth 30°. a. simulation, extended filaments; b. smoothed spectral-image plot, extended filaments. c. simulation, no extended filaments, d. smoothed spectral-image plot, no extended filaments. Panels e-h, same as top row for azimuth 130°